# On the material geometry of continuously defective corrugated graphene sheets

Andrzej Trzęsowski
Institute of Fundamental Technological Research
Polish Academy of Sciences
Pawińskiego 5B, 02-106 Warsaw, Poland
e-mail adresses: atrzes@ippt.pan.pl, artrzes@gmail.com

**Abstract.** Geometrical objects describing the material geometry of continuously defective graphene sheets are introduced and their compatibility conditions are formulated. Effective edge dislocations embedded in the Riemann-Cartan material space and defined by their scalar density and by local Burgers vectors , are considered. The case of secondary curvature-type defects created by this distribution of dislocations is analysed in terms of the material space. The variational geometry of the material space closely related with the existence of a characteristic length parameter is proposed. The formula which describes, in a reference temperature, the influence of dislocations on the material Riemannian metric, is given.

## 1. Introduction

Imperfections in the lattice of crystalline solids are responsible for many of the physical and chemical properties of these solids. However, it turns out that a crystal structure can be distorted only in a few, well-defined way. [11] For example, there exist two kinds of the so called *line defects*: dislocations and disclinations. *Dislocation* results from the translation, or linear displacement, of one part of the crystal with respect to another part. *Disclination* is formed when one part of a crystal structure is displaced by a rotation rather than by a translation. However, disclinations are seldom observed in ordinary three-dimensional crystals such as those of metals. They do appear, for example, in liquid crystals and they are important structural elements in many ordered materials other than conventional crystals, such as protein coats of viruses.[11]

The geometry of both single dislocations and disclinations can be illustrated with the aid to consider multiply-connected three dimensional objects being an imaginary solid bodies that material has no microscopic structure and has ideal mechanical properties. For example, it is considered – as a prototype of such an object – a hollow cylinder made out of elastic material and cut it at a radial half two-plane. It destroys its multiple connectedness. "Then it is taken the two lips that have been separated by the cat and translate and rotate them against each other. Finally, after eventually removing or adding missing material, the two planes are welded together again. This cutting and welding process is called the *Volterra process* (called also *Volterra's cut-and-weld operation*). The Volterra process, by construction, yields six different kinds of distorted objects, which belong to the six degrees of freedom of the *Euclidean group* E(3) of motions in three dimensions: the distortions belonging to the *translational* sub-group T(3) and to the *rotational* subgroup SO(3) are dislocations and disclinations, respectively." [30]

Notice, that the *topological classification* of single line defects of Bravais crystals with irremovable distortion, which is based on the consideration of the *group of affine motions* in a three-dimensional affine point space (see [39] – Appendix A), is consistent with the Volterra process but predicts a new type of line defects in crystal lattices, which is not reducible to dislocations or disclinations and represents general *shear distortion*.[31] A topological invariant which reflects the magnitude and direction of the crystalline lattice distortion produced by a dislocation is represented by a non-zero translation - the Burgers vector of this dislocation. Next, according to this topological classification of line defects, while in the *three-dimensional* case there exists exactly one type of irremovable distortion corresponding to the line defects of rotation type, in the *two-dimensional* case, the rotation type line defects with irremovable non-equivalent distortions are in one-to-one correspondence with the non-zero integers. [31]

The *Burgers vector* quantitatively characterizes *dislocations* and can be defined in the following simplified way (e.g. [5], [12]). Namely, it is considered the so-called *Burgers circuit* enclosing the end of the line cut (see Volterra process) with discrete lattice steps. If the dislocation occurs, then a vector being a proper translational vector of the lattice has to be added in order to close this circuit. This vector is the Burgers vector and its *strength* (i.e., its module) defines the value of lattice translational distortion. The Burgers vector is defined up to its orientation. Without dislocation the Burgers circuit will end up at the starting point and thus the Burgers vector vanishes. Dislocations in a three dimensional crystal can be classified by the direction of the Burgers vector with respect to the plane in which the Burgers circuit is located: the Burgers vector of an *edge dislocation* lies in this plane, the Burgers vector of a *screw dislocation* is perpendicular to this plane, and a *mixed dislocation* is partially edge and partially screw.[40] It follows from these definitions that in *planar crystals* (as graphene is) only edge dislocations exist.

A *disclination* is characterized by a closure failure of the rotation, called the *Frank angle*, for a closed circuit round the disclination line. There are wedge and twist disclinations [11]. The *wedge disclination* can be illustrated by adding a wedge of atoms ([5], [11]). Since a disclination line in a three-dimensional crystal breaks its rotational symmetry, it is generally represented by a rotation vector**,** called the *Frank vector* of the disclination and perpendicular to the plane of rotation. Its strength equals the Frank angle and defines the value of rotational distortion. However, since in three dimensional solid crystals disclinations needs high activation energy, this kind of line defects is not observed in such crystals ([17], [36]). The Frank vector cannot be defined in terms of the two-dimensional lattice only but the Frank angle can be still defined.

The symmetries of crystals impose constraints on dislocations and disclinations. For example, the only instances where the continuity of the crystal lattice can be preserved are those in which the rotation of the disclination is a symmetry rotation of the lattice. The Burgers vector is, by definition, a translation vector of the lattice. It is, for energetic reasons, usually the shortest translation vector of the lattice.[11]

Let us formulate, basing oneself on the papers [26], [44] and [45], some facts concerning defects of the graphene crystal structure. The properties of 2D solid materials can be strongly affected by structural irregularities. Graphene edges (there are two basic shapes for graphene edges, namely armchair and zig-zag edges) and point defects such as vacancies have to be distinguished from dislocations and grain boundaries, structural defects characterized by the finite values of their respective topological invariants, Burgers vectors and misorientation angles. Such topological defects as well as disclinations introduce the *non-local disorder* into the crystalline lattice.[44]

The honeycomb lattice of graphene has underlying Bravais lattice with basis vectors which can be chosen as [44]:

$$\mathbf{a}_1 = \frac{a}{2}\mathbf{i} + \frac{\sqrt{3}a}{2}\mathbf{j}, \qquad \mathbf{a}_2 = -\frac{a}{2}\mathbf{i} + \frac{\sqrt{3}a}{2}\mathbf{j},$$
$$a = \sqrt{3}d = 2.46\,\text{Å}, \quad d = 1.42\,\text{Å}, \quad \|\mathbf{a}_1\| = \|\mathbf{a}_2\| = a, \tag{1}$$

where $d$ is the nearest neighbour distance in graphene, $a$ is the nearest neighbour distance in a sublattice of graphene, $\mathbf{i}$ and $\mathbf{j}$ are base vectors parallel of Cartesian coordinates in $\mathbb{R}^2$ which cover with the zig-zag and armchair *high-symmetry directions* of the lattice, respectively. The lattice of the *discrete material geometry* of graphene is made up of two sub-lattices, *A* and *B*, where *A* atoms occupy Bravais lattice nodes, and *B* are shifted by $\boldsymbol{\delta} = (\mathbf{a}_1 + \mathbf{a}_2)/3$ [26]:

$$\mathbf{r}_A(m, n) = m\mathbf{a}_1 + n\mathbf{a}_2,$$
$$\mathbf{r}_B(m, n) = m\mathbf{a}_1 + n\mathbf{a}_2 + \boldsymbol{\delta}, \qquad m, n \in \mathbb{Z}, \tag{2}$$

where $\mathbb{Z}$ is the set of integers.

In two-dimensional materials, as it was mentioned above, only *edge dislocations* are possible. The Burgers vector $\mathbf{b}$ of the dislocation (defined up to its orientation), which defines the magnitude $b = \|\mathbf{b}\|$ (the strength) and direction of the crystalline lattice distortion produced by a dislocation, is located in the *material's plane*.[44] Moreover, by the definition, any Burgers vector $\mathbf{b}$ is a proper translational vector of graphene lattice, that is, it can be written in the form:

$$\mathbf{b} = m\mathbf{a}_1 + n\mathbf{a}_2, \qquad m, n \in \mathbb{Z}. \tag{3}$$

Note that "the $\mathbf{j} = (0, 1)$ inserts a semi-infinite strip of atoms along the armchair high-symmetry direction in graphene while its Burgers vector is oriented along the zigzag direction and it is the shortest Burger vector with the strength $b = \sqrt{3}d = 2.46$ Å….The $\mathbf{i} + \mathbf{j} = (1, 1)$ dislocation has a larger Burger vector with the strength $b = 3d = 4.23$ Å and inserts a semi-infinite strip of width $b$ along the zigzag direction of graphene."[44]

Notice also that in graphene one-dimensional chains of edge dislocations constitute tilt *grain boundaries* with mutual orientation of two crystalline domains described by the *misorientation angle* $\alpha \in (0, \pi/3)$.[44]

Topological defects can be formed also by replacing a hexagon by n-sides polygon. Particularly, a *pentagon* induces the positive curvature while a *heptagon* induces the negative curvature. In the literature these two regular polygons are frequently identified with positive or negative disclinations, respectively.[44] Then, a pair of complementary disclinations at short distances can be seen as a dislocation and, consequently, in [44] arbitrary dislocations and grain defects in graphene are described starting from disclinations as the elementary topological defects. It is stated, as a conclusion, that dislocations and grain boundaries are important intrinsic defects in graphene which may be used for engineering graphene based nanomaterials and functional devices.[44] However, this statement is independent from the above mentioned identification and it seems geometrically more sensible to consider pentagons and heptagons as the (positive or negative) *elementary topological curvature-type defects*. In this approach dislocations as well as disclinations are particular cases of the second kind of topological defects – the *line defects* (see overhead mentioned topological classification of line defects) and can be formed in graphene by the combination of these curvature-type topological defects.

Pentagons and heptagons generally can appear in pairs in the graphene structure since the mean coordination number of the plane trivalent polygonal cell systems equals 6 according to the Euler's law. There are two characteristic examples of topological defects build up from these two types of regular polygons; *Stone-Wales* defect and *mitosis* [45]. So, let us consider the *general Stone-Wales transformation* denoted as $SW_{p/r}$ and changing a group four proximal faces with $p$, $q$, $r$, $s$ atoms to four new rings with $p$-1, $q$+1, $r$-1, $s$+1 atoms. Namely, "$SW_{p/r}$ reversibly rotates the bond shared by the two rings $p$ and $r$, preserving both, the total number of carbon atoms $v$ and the total number $c$ of carbon-carbon bonds:

$$v = p+q+r+s-8, \qquad c = v+3. \qquad (4)$$

For example, on the *graphene ideal surface*, made only of hexagonal faces, the $SW_{6/6}$ rotation transforms for hexagons into 5|7 adjacent pairs symbolized as 5/7/7/5 defect also quoted as the SW defect or the *dislocation dipole*." [27] …"The *Stone-Wales* defect is the $\pi/2$ rotation of two carbon atoms with respect to the midpoint of the bond. In this defect four adjacent hexagons are changed into two pentagons and two heptagons (two pentagon-heptagon 5|7 pairs). The pentagons are separated by heptagons. The number of the atoms (vertexes) does not change… The arrangement containing one defect remained planar. Both the bond lengths and the angles becomes distorted."[45]

Several examples of the structures containing more pentagon-heptagon pairs but *preserving flatness* of the graphene sheet are known. For example a structure containing these defects arranged in a line or in a net remains planar.[45] The pentagon-heptagon 5|7 pairs iteratively propagated in the graphene layer produce a structural defect called *Stone-Wales wave*. [27]

The fact that the graphene structure containing Stone-Walles defects remain planar should be related with the fact that this defect can be seen as a dislocation dipole. It seems that there is a cancellation in the distortions originated by the dislocations [45] (cf. the identification of these polygonal defects proposed in [44]). Notice that a model of *amorphous graphene* can be generated by introducing Stone-Wales defects into perfect honeycomb lattice [13]. It is important from the point of view of physical applications of defective graphene sheets that, as it is stated in [5], "disclinations (formed by isolated pentagon or heptagon rings), dislocations (pentagon-heptagon pairs) and Stone-Walles defects (special dislocation dipoles) were found to have the least formation energy and activation barriers."

Let us quote now some statements formulated in the paper [45] and concerning the *mitosis*. „The mitosis is a lattice defect where two pentagons originate from a given hexagon and con-sequently the neighbouring hexagons become heptagons. The heptagons are separated by pentagons. The number of atoms increases by two. In this case the distorted graphene structure is a planar structure if the heptagon pair separated by the pentagon pair is studied alone, but the structure is not planar if this defect is constrained in the graphene structure. If the mitoses are arranged next to other along a line, the structure distortions are summarized along the line, the sum is very large, and the solutions is a *wavy pattern* with alternating curvatures. Mitosis can be arranged next to each other not only along straight lines but along curves or groups. For example: three pentagons placed next to each others produces a larger curvature in the graphene structure than in the case two pentagons. It is interesting that the largest curvature arises from six mitoses arranged in a group. In this case, six pentagons placed next to each other created a half dodecahedron, which can be the end of an armchair-type nanotube. If the six pentagons are arranged along a curve, the resulting structure is the end of a zigzag-type nanotube. The arrange more than six pentagons next to each other cannot be solved in a pentagon-heptagon-hexagon system. So, the pentagons and/or heptagons can occur in the system alone. When a pentagon is surrounded by hexagons, a *spherical surface* forms and when a heptago*n* is surrounded by hexagons, the characteristic *saddle-shaped surface* forms."

To study the *stability* of the distorted structures, the cohesive energy (the average energy of the chemical bonds: the total energy divided by the number of the bonds) was calculated in [45] for the above discussed structures. Cohesive energy in the environment of the defects increases several percent compared to the cohesive of the perfect graphene in every case. The increase is smaller for the planar structures and it is larger for the structures with curvatures. The more pentagons are connected to each other, the larger the decrease of stability. The mitoses arranged along a straight line have the least stability because the pentagons cause curvatures in both sides of the graphene sheet where they are connected with each other.

So, stable dislocations in the graphene lattice can be made of pentagon-heptagon pairs. These dislocations are called *glide dislocations*. Notice that there exist also stable dislocations called *shuffle dislocations* and made by an octagon with dangling bond. A shuffle dislocation contains an additional atom with respect to the glide dislocation. Every of these two kinds of edge-type line defects add one or several lines of atoms to the graphene lattice.[5]

Although many proposed applications of graphene require the existence of graphene endowed with an extended one-dimensional defect, nevertheless the existence of graphene sheets endowed with line defects is, in fact, the matter of nanomaterial engineering. It is because "the energy cost for formation of extended defects in covalently bonded materials is high. This makes the spontaneous formation of such one-dimensional defect during the growth of graphene highly unlikely."[24] For example, experiments have revealed that whereas mechanically exfoliated monolayer graphene sheet is structurally almost perfect in atomic scale [27], the growth of single- or few-layer graphene on Cu and Ni substrates through chemical vapour deposition produces polycrystalline graphene sheets with many grain boundaries consisted of a continuous array of pentagon/heptagon pairs [25]. The electron mobility in these graphene sheets are significantly different for those of pristine graphene.[25] Note also that because in graphene the dynamics of lattice defects occurs at a much higher energy than the electronic processes [5], the motion of electrons (observed in the three-dimensional configurational space Minkowski type – [39]) can be considered in the time-independent geometries of defective graphene sheets (called *frozen geometry* in [5]).

The paper is organized as follows. In Section 2 crystalline defects in two and three-dimensional continuized defective crystals are described and their different geometrical representations are presented and discussed. In Section 3 compatibility conditions of material geometrical objects describing continuously defective crystals are discussed and the notion of effective edge dislocation lines is introduced. In Section 4 the problem of consistent parametrization of metric and non-metric geodesics in the material space of continuously defective graphene sheets is discussed. In Section 5 continuously dislocated corrugated graphene sheets with the flat material Riemann-Cartan geometry, are considered. The notion of secondary curvature-type defects created by the distribution of dislocations is introduced and analysed. In Section 6 the Riemann-Cartan variational geometry of the material space closely related with the existence of a characteristic length parameter, is proposed.

**2. Material geometry of continuously defective crystals**

Let us start with the quoting of the way in which the notion of continuized defective crystals is introduced in the paper [19]. "Since the crystal is a discrete assembly of particles, a discrete geometry should be appropriate to describe the geometric aspects. However, the scale of discreteness is rather small compared to more macroscopic scales that are important in many applications. It is then justified to neglect the finiteness of the atomic spacing in the crystal and to introduce the concept of "*continuized crystals*" by a thought limiting process: The "atoms" are reduced again and again to smaller size and are arranged in a lattice of the same type, however with smaller lattice constants. At the same time, also the defects have to be cut to pieces and distributed in the new lattice such that the content of defects, like that of mass, per macro-volume element remains unchanged. Thus the continuized crystal is the true image of the real crystal."

Let us quote also an "engineering" (slightly modified) point of view concerning the physical meaning of the notion of continuized three-dimensional Bravais crystals taken from [40] (see also the references therein). "Let us take 1mm as a macroscopic observation level and let 1Å (the diameter of the hydrogen atom in the ground state) define the atomic-size observation level scale. It is known, for example, that for usual well-annealed pure metals, the mean distance between dislocations is of the order 1μm (1μm = $10^{-3}$ mm), and that the crystal with many dislocations can be treated, on mesoscopic observation level scale that lies e.g. in the range 10-100 nm (1nm = $10^{-3}$ μm = 10 Å), as a part of an ideal crystal". Notice that 10 nm ≈7$d$ where $d$ is the nearest neighbour distance in graphene; $d$ equals the radius of the circle circumscribed on the elementary hexagon of the graphene lattice…."On the other hand, if the macroscopic properties of a crystalline solid with many dislocations are considered, a *continuous limit approximation* can be defined by means of the condition that, at each point of the material body, a *characteristic mesoscopic length*, say of the order of 10-100 nm, can be approximately replaced with the *infinitesimal length*. Consequently, a monocrystal with *many dislocations* can be considered as such *locally homogeneous* continuous body that retains locally the most characteristic properties of the original crystal, namely the existence of *crystal-* *lographic directions* at each point, the *lattice rotational symmetries* (the lattice translational symmetries are lost in a continuum limit) and the mass density. Therefore, although the global long-range order of crystals is lost in the presence of dislocations, nevertheless their local long-range order still exists. It is represented by the *object of material anholonomity* of the continuized dislocated crystal."[40] Note that the content of defects, e.g. the *scalar density of dislocations* defined in the case graphene sheets as the number of all dislocation lines in surface unit (Section 3), remains unchanged in this continuous limit.

It ought to be stressed that in *classical crystals* (i.e., possessing as local degrees of freedom the translational degrees of freedom only) "this limiting process is possible for dislocations, but not for disclinations, since these defects are specified by a finite angle (symmetry of the crystal) which must not be divided into smaller angles in the limiting process. This means that a crystal with disclinations cannot be continuized in Euclidean, or, more generally, in flat space (space without curvature)"([18]; see also [22]). Therefore, "disclinations in solids (not in liquid !) crystals can be viewed as sources of curvature (in the sense of differential geometry) within the otherwise flat crystal, or they are themselves nothing else than such curvatures. It is then suggestive to speak of flat crystals or curved crystals. In this terminology the only elementary (i.e., not composed) line defect in flat classical crystals is the dislocation."[18] Notice that it is not, for example, the case of *Cosserat* crystalline solids ([21], [22]). "Obviously, curved (classical) crystals are possible only if the curvature is, in some sense, compatible with the considered crystal structure. One may image defects in curved crystals. However, the curvature itself is not a defect. A curved crystal situation with continuous curvature can be created, for instance, by nonuniform heating up of a crystal"[18]. Moreover, in the case of *corrugated graphene sheets*, the occurrence of a correlation of the surface corrugations with its thermal state is observed.[39]

The appearance of dislocations in an initially planar graphene sheet can leads to a bend of originally straight lattice lines and consequently planes spanned by two crystallographic directions become *local crystal planes*. If, additionally, local crystal planes of a graphene sheet are *virtually local slip planes* (i.e., local planes in which infinitesimal dislocation lines can move), the graphene sheet is virtually a *slip surface* along which a curved edge dislocation line can move. In general, it is known that the glide motion of many dislocations results in slip, and it is observed that globally (i.e., on a *macroscopic scale*) this motion is accompanied by the occurrence of slip surface (this is the considered graphene sheet in our case) in which *dislocations* can move. Such crystal surfaces (graphene sheets in our case) are called *glide surfaces* of dislocations. Particularly, in the case of *single glide* a planar graphene sheet pass into a slip surface without local stretchings.[41] In this case the graphene sheet becomes a *developable* glide surface (see [39]).

The *slip*, which is the most common manifestation of *plastic deformation* in crystalline solids, can be envisaged as successive displacement of one part of crystal with respect to another. In a three-dimensional crystal, it is a displacement of one plane of atoms over another, on a *distinguished slip plane* (local or global). In a graphene sheet the slip plane is tangent to this crystalline surface and the displacement of atoms (homogeneous or not) is defined by a direction located in this plane and normal to a line located on the sheet. "Consequently, any *dislocation line* in the crystal can be treated as a line formed by means of a slip (homogeneous or not), such that the dislocation becomes a boundary between slipped an unslipped parts of the crystal. The *slip direction* is then parallel to the Burgers vector of the dislocation, and the *slip magnitude* equals the strength of the Burgers vector (defined as the modulus of its Burgers vector). This representation of a dislocation concerns straight as well as curvilinear dislocation lines and the dislocations so represented are called *Volterra dislocations*" ([41] and references therein; see also Section 1 of this paper).

Thus, in a continuized crystal (the two-dimensional crystal in our case), the line being a boundary between slipped and unslipped parts of the crystal and located on the glide surface can be distinguished. This line can be endowed, in the framework of the geometrical theory of dislocations, with the so-called *local Burgers vector* (Section 5; see also [40]) tangent to the slip surface along the line everywhere. The line endowed with a local Burgers vector is called a *Volterra-type effective dislocation line*.[41] It ought to be stressed that the local Burgers vector is not constrained by the condition (3) and the *glide motion* of a Volterra-type effective dislocation can be considered as a *mesoscopic elementary act* of macroplasticity in a continuized crystal.[41] Moreover, in flat surfaces *screw dislocations* do not exist (see Section 1). Since the corrugated graphene sheets considered in the paper are homeomorphic to the developable surfaces [39], dislocations in graphene can be represented, in the continuous limit approximation, by the *Volterra-type effective edge dislocations* which are defined as a smooth curve located on the such sheet and endowed with a local counterpart of the Burgers vector being a nonvanishing everywhere smooth vector field tangent to the sheet and normal to the curve (see Section 3 and [41]).

Let us define, as an auxiliary notion, a local infinitesimal counterpart of the Burgers vector. A standard simplified way in which this auxiliary notion can be defined is the following (see [40] for the more precise reasoning). Dislocations represent, from the geometrical point of view, the source of translational mismatch ("*anholonomy*") and distributions of many dislocations can be described, in the continuous limit approximation, with the help of differential geometry (e.g. [17] – [19], [21], [40] – [42]). Firstly, in differential geometry, if

$$\mathbf{v}(u) = v^\alpha(u)\partial_\alpha,$$
$$\partial_\alpha = \partial/\partial u^\beta, \qquad u = (u^\alpha),$$
(5)

is the local coordinate description of a vector field **v** tangent to a differential manifold *M* endowed with the covariant derivative $\nabla$ defined by the Christoffel symbols $\Gamma^\kappa_{\alpha\beta}$:

$$\nabla_\alpha \partial_\beta = \Gamma^\kappa_{\alpha\beta}\partial_\kappa,$$
(6)

(e.g. [4], [32] and [34]; see also Appendix A – remarks following eq.(155)), then the vector **v**(*u*) tangent to *M* at a point possessing the coordinates *u* is *parallel transported* from *u* to *u* + d*u*, if

$$\delta v^\alpha := \mathrm{d}u^\kappa \nabla_\kappa v^\alpha = \mathrm{d}v^\kappa + \Gamma^\kappa_{\alpha\beta}v^\alpha \mathrm{d}u^\beta = 0,$$
(7)

where d*u* = (d*u*$^\alpha$) describes an infinitesimal change of coordinates. It is equivalent to the statement that the coordinates $v^\alpha$ of the vector **v**(*u*) experience infinitesimal variations d$v^\alpha$ given by

$$\mathrm{d}v^\kappa = -\Gamma^\kappa_{\alpha\beta}v^\alpha \mathrm{d}u^\beta.$$
(8)

Next, in order to formulate an equivalent of the *Burgers circuit* from the definition of the Burgers vector (Section 1), we will consider the "infinitesimal vector" $d\mathbf{v}(u)$ and the following approximate formula (see [32]) describing the change of the vector field $\mathbf{v}$ due to its parallel transport defined by the infinitesimal change of coordinates $u$ to $u + du$:

$$\mathbf{v}(u + du) \simeq \mathbf{v}(u) + d\mathbf{v}(u),$$
$$d\mathbf{v}(u) := dv^\alpha(u)\partial_\alpha. \tag{9}$$

Let us denote

$$\mathbf{v}_a(u) := d_a u^\kappa \partial_\kappa, \quad u = (u^\kappa),$$
$$d_a u = (d_a u^\kappa), \quad a = 1, 2. \tag{10}$$

If two infinitesimal vectors $\mathbf{v}_1(u)$ and $\mathbf{v}_2(u)$ are displaced parallel along each other, the figure obtained is not parallelogram but a *pentagon* with the closing vector:

$$\delta\mathbf{b}(u) := \mathbf{v}_2(u + d_1 u) - \mathbf{v}_1(u + d_2 u) \simeq \delta b^\kappa(u)\partial_\kappa,$$
$$\delta b^\kappa := \varepsilon S_{\alpha\beta}{}^\kappa df^{\alpha\beta}, \quad S_{\alpha\beta}{}^\kappa = \Gamma^\kappa_{[\alpha\beta]} = \frac{1}{2}\left(\Gamma^\kappa_{\alpha\beta} - \Gamma^\kappa_{\beta\alpha}\right), \tag{11}$$

where $S_{\alpha\beta}{}^\kappa$ are components of the so-called *torsion tensor* (Appendix), the multiplier $\varepsilon = \pm 1$ appears because the infinitesimal vector $\delta\mathbf{b}$ is defined up to the choice of its orientation, and

$$df^{\alpha\beta} = 2 d_1 u^{[\alpha} d_2 u^{\beta]} \tag{12}$$

is the infinitesimal contravariant bivector at $u$.

Let us assume that an infinitesimal contravariant bivector $df^{\alpha\beta}$ at $u$ of the form of eq.(12) is represented by a part of an affine two-dimensional point space $A_2$ (with inner orientation) with a definitely fixed boundary curve passing through $u$. This bivector can be interpreted then as the *oriented area element*.[32] Next, let us consider now the parallel displacement of a vector $\mathbf{v}(u)$ along this infinitesimal curve, and let this displacement start and end at $u$. It can be proved that the difference $d\mathbf{v}(u)$ between the final and the initial value can be written in the form

$$d\mathbf{v}(u) = dv^\kappa(u)\partial_\kappa, \quad dv^\kappa = -\frac{1}{2} R_{\sigma\mu\lambda}{}^\kappa v^\lambda df^{\sigma\mu},$$
$$\frac{1}{2} R_{\sigma\mu\lambda}{}^\kappa = \partial_{[\sigma}\Gamma^\kappa_{\mu]\lambda} + \Gamma^\kappa_{[\sigma|\rho|}\Gamma^\rho_{\mu]\lambda}, \tag{13}$$

where $R_{\sigma\mu\lambda}{}^\kappa$ are components of the so-called *curvature tensor* of the covariant derivative $\nabla$ defined by the connection coefficients $\Gamma^\kappa_{\alpha\beta}$ (Appendix). Notice that the difference $d\mathbf{v}(u)$ lies in the plane of $df^{\sigma\mu}$ if and only if (in designations of [32] and [34]):

$$(n-1) R_{\alpha\mu\lambda}{}^\kappa = 2\delta^\kappa_{[\alpha} R_{\mu]\lambda}, \quad R_{\mu\lambda} := R_{\kappa\mu\lambda}{}^\kappa, \tag{14}$$

where $R_{\mu\lambda}$ is called the *Ricci tensor*.[32]

**Statement 1** *If n = 2, that is we are dealing with a continuously defective corrugated graphene sheet, then the general form of the curvature tensor is given by eq.(14). It means that in this case the infinitesimal difference between the final and initial position of a parallel displacement of a vector along the boundary of an infinitesimal parallelogram $df^{\beta\lambda}$ lies in the plane tangent to this graphene sheet. Consequently, this difference can be interpreted as an infinitesimal measure of the continuous counterpart of discrete disclinations in a graphene sheet.*

We see that the above procedures of parallel displacement of a vector field assign to an area element a small translation and a small rotation.([12], [21], [32]) It is similar to what one does in gauge field theory where a global spatial symmetry (defined e.g. by Euclidean motions of translations and rotations) is extended to a local material symmetry [12]. In the paper [38] this approach is applied to the case of continuous distributions of dislocations and, as a consequence, the tensor of *internal couple stresses* analogous to the tensor of couple stresses considered in the theory of polar continua but caused by the self-interaction of dislocations appears. However, it ought to be stressed that the same kind of internal stresses in continuously dislocated bodies appears also in the field theory based on the global invariance (under the proper full linear group $GL^+(3)$) proposed in [37]. It suggests the following presumption consistent with the curvature effects discussed in [39] (see also [21], [10], [12]):

**Hypothesis 1** *The internal forces in corrugated graphene sheets with a continuous distribution of line defects can be estimated in the framework of the theory Cosserat surfaces.*

The appearance of the oriented area element means that, in the description of continuously defective graphene sheets, we are dealing with a two-dimensional affine manifold $(M, \nabla)$ and additionally with a two-dimensional "volume-form", i.e., with an *oriented area form*. It suggests to consider the *material geometry of graphene sheets* defined by the triple $(M, \nabla, \omega)$ where $\omega$ is a general two-form. The two-form can be constructed on every open subset $O$ of $M$ by using a moving coframe $e^* = (e^a, a = 1,2)$ consisting with two linearly independent 1-forms and a positive scalar $f$:

$$\omega = f\, e^1 \wedge e^2, \quad f \in C^\infty(M),$$
$$e^1 \wedge e^2 = e^1 \otimes e^2 - e^2 \otimes e^1. \tag{15}$$

Let $e = (\mathbf{e}_a, a = 1,2)$ be a moving frame of smooth vector fields tangent to $M$, defined on the set $O$, and such that $e^*$ is the moving coframe dual to $e$. Consequently, we are dealing, at least locally, with a *distinguished moving frame* $e = (\mathbf{e}_a, a = 1,2)$ of smooth vector fields tangent to $M$ whose orientation is represented by the corresponding area form. It is a continuous counterpart of the distinguished lattice vectors of graphene considered in Section 1. Particularly, we say that $(M, \nabla)$, $\dim M = n$ ($n = 2$ in our case) is *locally equiaffine* (or *volume preserving*) if locally, around each point of $M$, there exists a nonvanishing and covariantly constant $n$-form $\omega$:

$$\nabla \omega = 0. \tag{16}$$

In this case ω is called a (local) *volume element* [43]. Notice that frequently (e.g. [16]) the volume element on *M* is defined as a nonvanishing *n*-form; in the paper we will call such *n*-form – a *volume form* (a *surface form* if *n* = 2).

Let us consider the case when the continuous description of material properties of a crystalline body needs to introduce, except the covariant derivative ∇, a Riemannian metric (see [39] in the case of graphene sheets or, e.g., [17], [38], [40] - [42] in the case of three dimensional continuized Bravais crystals, and [21] in the case of a Cosserat three-dimensional crystal):

$$\mathbf{a} = a_{\alpha\beta} \mathrm{d}u^\alpha \otimes \mathrm{d}u^\beta . \tag{17}$$

In this case we can introduce the *internal* (*"material"*) *length measurement* in the considered continuously defective body:

$$v_a \equiv \|\mathbf{v}\|_a = (\mathbf{v},\mathbf{v})_a^{1/2} = \sqrt{a_{\alpha\beta} v^\alpha v^\beta} , \tag{18}$$

where eq.(5) was taken into account and the differential operator $\partial_\alpha$ defines the *operational representation* of the vector field tangent to the $u^\alpha$ – coordinate curve (see [40] – Appendix). Next, let us consider a dimensional curvilinear coordinate system $u = (u^\alpha)$ such that $[u^\alpha]$ = cm (see [39] and [40]). Then $[\mathrm{d}u^\alpha]$ = cm, $[\partial_\alpha := \partial/\partial u^\alpha]$ = cm$^{-1}$, and, for example, if $\mathbf{v} = v^\alpha \partial_\alpha$, $[\mathbf{v}]$ = cm$^{-1}$ is a tangent vector field, then $[v^\alpha]$ = [1].

It follows from eqs.(5)-(8) that the parallel transport of a vector field **v,** [**v**] = cm$^{-1}$, preserves its dimension only if

$$\left[ \Gamma^\kappa_{\alpha\beta} \right] = \mathrm{cm}^{-1} . \tag{19}$$

Next, if we will assume that the base vector fields $\mathbf{e}_a$ have the dimension cm$^{-1}$, then $[e^a]$ = cm, [ω] = cm$^2$, and it follows from eqs.(11)-(19), that

$$\begin{gathered}
\left[ \delta \mathbf{b} \right] = [1], \quad \left[ \mathbf{a} \right] = \mathrm{cm}^2, \quad \left[ \delta b^\kappa \right] = \mathrm{cm}, \\
\left[ a_{\alpha\beta} \right] = [1], \quad \left[ R_{\sigma\mu\lambda}{}^\kappa \right] = \left[ R_{\mu\lambda} \right] = \mathrm{cm}^{-2}.
\end{gathered} \tag{20}$$

## 3. Compatibility conditions of material geometrical objects

The real graphene sheet is always embedded in a configurational orthogonal space (Euclidean or Minkowski type) and is homeomorphic to an ideal graphene sheet (defined as subset of a two-dimensional Euclidean point space homeomorphic to the disk in this space).[39] A real graphene sheet can be endowed with the *ideal material geometry* only when this sheet has a planar configuration. The material geometry can be not ideal but *perfect* in this sense that it is generated from the ideal ones by its distortion which does not introduce defects into the crystal structure of graphene. For example, a perfect crystal can be realized by a bent of a planar configuration of the graphene sheet endowed with the ideal material geometry such that the *flatness* of this configuration is preserved. These are *developable* graphene sheets. However, it ought to be stressed that there are such kinds of defects of the crystal structure of graphene sheets which produces developable configurations of these sheets ([39]; see Section 1). There are also not flat *Riemannian perfect material geometries* of graphene sheets. A physical realization of configurations endowed with such a material geometry is due to the existence of a varying temperature field. It can be, for example, a temperature field strictly correlated with the observed existence of corrugations of the two-dimensional ideal graphene crystal structure in the three-dimensional Euclidean configurational space [39]. Moreover, a varying magnetic field can produce a not flat Riemannian perfect graphene sheet and then such sheet is frequently considered as a space-like surface embedded in the three-dimensional Minkowski space [39].

In the paper we are dealing with various geometrical objects called *material* and introduced in order to describe the corrugated graphene sheets with their continuized crystalline structure *inelastically distorted* due to the existence of lattice defects. Consequently, the problem of compatibility conditions of these geometrical objects arrives. In order to discuss this problem, we briefly recall some definitions. The components $S^{\kappa}_{\alpha\beta}$ of the torsion tensor of eq.(11) are defined (in terms of the absolute notation of geometrical objects) by

$$S(\mathbf{r}, \mathbf{s}) = \frac{1}{2} T(\mathbf{r}, \mathbf{s}) = \left(S_{\alpha\beta}{}^{\kappa} r^{\alpha} s^{\beta}\right) \partial_{\kappa} , \qquad (21)$$
$$\mathbf{r} = r^{\alpha} \partial_{\alpha}, \quad \mathbf{s} = s^{\beta} \partial_{\beta} ,$$

where [4]

$$T(\mathbf{r}, \mathbf{s}) := \nabla_{\mathbf{r}} \mathbf{s} - \nabla_{\mathbf{s}} \mathbf{r} - [\mathbf{r}, \mathbf{s}], \qquad (22)$$
$$[\mathbf{r}, \mathbf{s}] := \mathbf{r} \circ \mathbf{s} - \mathbf{s} \circ \mathbf{r}.$$

If, as it is in eqs.(10) and (11), the vector fields $\mathbf{r}$ and $\mathbf{s}$ are displaced $\nabla$-parallel along each other, that is

$$\nabla_{\mathbf{r}} \mathbf{s} = \nabla_{\mathbf{s}} \mathbf{r} = 0 , \qquad (23)$$

then, according to eq.(22), we can define the following vector field $\mathbf{B}$ indirectly dependent on the covariant derivative $\nabla$:

$$\mathbf{B} := \varepsilon T(\mathbf{r}, \mathbf{s}) = -\varepsilon [\mathbf{r}, \mathbf{s}], \quad [\mathbf{B}] = \text{cm}^2 , \qquad (24)$$

where $\varepsilon = \pm 1$ defines the choice of the orientation of this vector field. We define, following the approach formulated in the papers [40] – [42] but in a modified form adapted to the considered two-dimensional manifolds, the following local (but not infinitesimal) counterpart of the local infinitesimal Burgers vector introduced in Section 2.

**Definition 1** *If the vector fields* $\mathbf{r}, \mathbf{s} \in W(M)$ *are linearly independent, then the formulae*

$$\mathbf{B} := \rho \mathbf{b} = -\varepsilon [\mathbf{r}, \mathbf{s}], \quad 0 < b_a = \|\mathbf{b}\|_a < \infty, \qquad (25)$$
$$\mathbf{b} = b^{\kappa} \partial_{\kappa}, \quad [b_a] = [b^{\kappa}] = \text{cm}, \quad [\rho] = \text{cm}^{-2},$$

*where eqs.(17) – (20) were taken into account, define the dimensionless local Burgers vector* $\mathbf{b}$ *associated with these vector fields and define the smooth non-negative scalar density of line defects* $\rho \in C^{\infty}(M)$. *This scalar density is defined (due the requirement of the consistency of dimensions) as the density of the total number of line defects located in the considered continuously defective graphene sheet and constrained by following conditions:*

$$\rho T(\mathbf{r}, \mathbf{s}) \neq 0, \quad 0 < N = \int_M \rho \omega < \infty , \qquad (26)$$

*where* ω *is a distinguished oriented surface form.*

Thus, according to this definition, we have to introduce a certain distinguished oriented *surface form* ω. This surface form can be defined by a metric tensor of the internal length measurement, by a distinguished moving frame or, for some physical reasons (see e.g. eq.(16) and remarks below eq.(39)), ω can be taken in the general form defined by eq.(15). The finite and positive scalar $b_a$ is called the *mean strength of dislocations* and its calculation needs to distinguish the metric tensor **a** which define an internal length measurement.

Let $e = (\mathbf{e}_a, a = 1,\ldots,n)$ be a distinguished base of the linear module W(*M*) of vector fields on *M* tangent to *M* and

$$\mathrm{e}^a = e^a{}_\kappa \mathrm{d}u^\kappa, \qquad \mathbf{e}_a = e_a{}^\kappa \partial_\kappa,$$
$$\mathrm{e}^a(\mathbf{e}_b) = e^a{}_\kappa e_b{}^\kappa = \delta^a_b, \qquad e = \det(e_a{}^\kappa), \qquad (27)$$

where $e = (e^a; a = 1,\ldots,n)$ is the base of the linear module of 1-forms W(*M*)* on *M* dual W(*M*) (see [40] – Appendix and Appendix here). Let us assume that for $n = 2$

$$N = \int_M \rho\omega = \int_M \rho \mathrm{d}F, \qquad \mathrm{d}F = e^{-1}\mathrm{d}u^1 \mathrm{d}u^2,$$
$$\omega = \mathrm{e}^1 \wedge \mathrm{e}^2 = e^{-1}\mathrm{d}u^1 \wedge \mathrm{d}u^2, \qquad \rho T_a \neq 0, \qquad T_a = \|T(\mathbf{e}_1,\mathbf{e}_2)\|_a. \qquad (28)$$

The condition (23) takes now the form equivalent to the condition (165), that is, we have:

$$\nabla_{\mathbf{e}_1}\mathbf{e}_2 = \nabla_{\mathbf{e}_2}\mathbf{e}_1 = 0. \qquad (29)$$

Moreover, according to eqs.(24) – (26) and (156), we obtain the following *local continuous counterpart* of eq.(3) associated with the base *e*:

$$\rho\mathbf{b} = [\mathbf{e}_1, \mathbf{e}_2] = C^c_{12}\mathbf{e}_c, \qquad b_a = \|\mathbf{b}\|_a > 0. \qquad (30)$$

It ought to be stressed that although the above definition of the local Burgers vector is not incompatible with the procedure of parallel displacement of vectors (Section 2 and the condition (23)) but does not take into account the form of the torsion tensor directly (the relation (24) is not assumed). The condition (29) can be treated then as the definition of a covariant derivative for which the condition (24) holds. In this sense, it is a generalization of the standard procedure of the definition of the local infinitesimal Burgers vector as well as a generalization of the discrete case. Thus, the local Burgers vector is uniquely defined by the pair (*e*, *ρ*) where *e* is a distinguished base of W(*M*) and *ρ* is the scalar density of dislocations. The surface form ω is defined then up to a positive scalar of eq.(15) (say, for example, defined by eq.(28)). The local Burgers vector defined in this way will be called *associated* with the moving frame *e*. The components $C^a_{bc}$ of eq.(30) (or of eq.(156) in the general case) defines the so called *object of material anholonomity* ([40]; see Section 2). In the paper we will consider *effective dislocation lines* understood as curved Volterra-type effective edge dislocation lines (Section 2) embedded in the material Riemannian space $M_a = (M, \mathbf{a})$:

**Definition 2** *Let $M_a = (M, \mathbf{a})$ be a two-dimensional Riemannian material space, $e = (\mathbf{e}_a) - a$ base of W(M), **b** – the associated local Burgers vector defined by eqs.(28) and (30). The congruence C[**l**] of lines in $M_a$ such that*

$$(\mathbf{l}, \mathbf{l})_a = 1, \qquad (\mathbf{l}, \mathbf{b})_a = 0, \qquad (31)$$

*defines a family of virtual Volterra-type effective edge dislocation lines of a continuized corrugated graphene sheet.*

So, let us consider a differential manifold $M$, $\dim M = n$, endowed with the quadruple ($\nabla$, **a**, $e$, $\omega$) of geometric objects introduced in order to describe continuously defective corrugated graphene sheets. The question arrives concerning the conditions of compatibility of these objects. Firstly, let us define (in a coordinate system) the tensor field **Q** of *nonmetric Christoffel symbols* $Q_{\alpha\beta}{}^{\kappa}$ (called also the *difference tensor*) being a part of the $\nabla$-Christoffel symbols $\Gamma^{\kappa}_{\alpha\beta}$ which is not obtainable from the metric (see Appendix):

$$Q_{\alpha\beta}{}^{\kappa} := \Gamma^{\kappa}_{\alpha\beta} - \tilde{\Gamma}^{\kappa}_{\alpha\beta}, \tag{32}$$

where

$$\tilde{\Gamma}^{\kappa}_{\alpha\beta} = \frac{1}{2} a^{\kappa\sigma} \left( \partial_{\alpha} a_{\beta\sigma} + \partial_{\beta} a_{\alpha\sigma} - \partial_{\sigma} a_{\alpha\beta} \right) \tag{33}$$

are metric Christoffel symbols of the Levi-Civita covariant derivative $\nabla^a$ corresponding to the Riemannian metric **a**. The components $S_{\alpha\beta}{}^{\kappa}$ of the *torsion tensor* **S** of the covariant derivative $\nabla$ defines the formula (see Appendix):

$$S_{\alpha\beta}{}^{\kappa} = \frac{1}{2} T_{\alpha\beta}{}^{\kappa} = \Gamma^{\kappa}_{[\alpha\beta]} = Q_{[\alpha\beta]}{}^{\kappa}, \tag{34}$$

and the so-called *non-metricity tensor* **W** ([3], [32]) is defined by:

$$W_{\alpha\beta}{}^{\kappa} := -a^{\kappa\sigma} \nabla_{\alpha} a_{\beta\sigma}. \tag{35}$$

The non-metricity tensor can be decomposed into a traceless part (*shear*) **P** and a trace (*Weyl covector*) $w = w_{\alpha} du^{\alpha}$ [3]:

$$W_{\alpha\nu\sigma} = a_{\sigma\kappa} W_{\alpha\nu}{}^{\kappa} = P_{\alpha\nu\sigma} - a_{\nu\sigma} w_{\alpha},$$
$$P_{\alpha\kappa}{}^{\kappa} = 0, \qquad w_{\alpha} = -\frac{1}{n} W_{\alpha\nu}{}^{\nu}. \tag{36}$$

The difference tensor **Q** can be decomposed into the so-called *contorsion tensor* **K**, expressing the difference between the Christoffel symbols of a metric-compatible part of the covariant derivative $\nabla$ and the Levi-Civita covariant derivative $\nabla^a$ and defined by [23]:

$$K_{\alpha\nu\sigma} = (S_{\alpha\nu\sigma} - S_{\alpha\sigma\nu} - S_{\nu\sigma\alpha}), \qquad S_{\alpha\nu\sigma} = a_{\sigma\kappa} S_{\alpha\nu}{}^{\kappa}, \tag{37}$$

where eq.(34) was taken into account, and a tensorial measure **V** of non-metricity:

$$V_{\alpha\nu\sigma} = \frac{1}{2}(W_{\alpha\nu\sigma} + W_{\nu\sigma\alpha} - W_{\sigma\alpha\nu}). \tag{38}$$

Namely, we have [23]:

$$Q_{\alpha\nu\sigma} = a_{\sigma\kappa} Q_{\alpha\nu}{}^{\kappa} = K_{\alpha\nu\sigma} + V_{\alpha\nu\sigma}. \tag{39}$$

Definition 1 introduces a distinguished surface form. Let us consider, basing oneself o the papers [23] and [33], the problem of compatibility of the covariant derivative $\nabla$ and the general volume element $\omega$. We are dealing, as it was mentioned at the beginning of this section, with the inelastically distorted graphene sheets. It is known that, in three-dimensional crystals, a special type of inelasticity being volume preserving elastoplasticity due to the glide motion of dislocations at sufficiently low temperatures (e.g. in room temperatures) occurs. It ought to be stressed that the constraint of volume preserving applies to the *plastic* (glide) motion of dislocations only [20] (see remarks in Section 2 concerning the glide motion). So, it is the case when both the physically distinguished volume or surface form ω and the *locally equi-affine* (eq.(16)) covariant derivative ∇ ought to be taken into account. The existence of a distinguished volume (surface) form enables also to define, for example, the *generalized divergence operator* $\text{div}_\omega$ by the rule [33]:

$$(\text{div}_\omega \mathbf{w})\omega := L_\mathbf{w}\omega, \quad \mathbf{w} \in W(M), \tag{40}$$

where $L$ denotes the Lie derivative operator (e.g. [1]) and $\dim M = n$. It, in turn, can be essential if the defective crystals theory is developed as a field theory consistent with the material geometry described by the triple (∇, ω, **a**) of geometrical objects. An example of using volume form compatible with a covariant derivative for minimal action principle is discussed in [23] and [33].

If **a** is the metric tensor of eq.(17) and a base $e = (\mathbf{e}_a)$ of the linear module W(M) fulfils the **a**-orthonormality condition

$$(\mathbf{e}_a, \mathbf{e}_b)_a = \delta_{ab}, \tag{41}$$

then we can define the natural *volume element* $\omega_a$ (see Section 2) on the Riemannian manifold $M_a = (M, \mathbf{a})$ by:

$$\omega_a(\mathbf{e}_1, \ldots, \mathbf{e}_n) = 1, \quad n = \dim M, \tag{42}$$

and the definition is independent of the choice of the base $e$. So, in this case [4]:

$$\omega = \omega_a = \sqrt{a}\, du^1 \wedge \ldots \wedge du^n, \quad a = \det(a_{\alpha\beta}), \tag{43}$$

and

$$\text{div}_{\omega_a}\mathbf{w} \equiv \text{div}_a \mathbf{w} = a^{-1/2}\partial_\alpha(a^{1/2}w^\alpha), \tag{44}$$

where eq.(17) was taken into account. It follows from eqs.(27) and (41) – (43) that

$$e^{-1} = \sqrt{a}. \tag{45}$$

Let us introduce designations:

$$\begin{aligned} \mathrm{t} &= t_\sigma du^\sigma, & t_\sigma &= T_{\alpha\sigma}{}^\alpha, \\ \mathrm{v} &= v_\sigma du^\sigma, & v_\sigma &= V_{\alpha\sigma}{}^\alpha, \end{aligned} \tag{46}$$

and let us take into account that if the *n*-dimensional manifold $M$ is endowed with a metric **a,** an arbitrary *volume form* ω can be written as

$$\omega = f\, \omega_a, \quad f \in C^\infty(M), \tag{47}$$

where $f$ is a positive function. Notice that if $e = (\mathbf{e}_a; a = 1,\ldots,n)$ is a local **a**-*orthonormal* basis of W(M) and $e^* = (\mathrm{e}^a; a = 1,\ldots,n)$ is the basis of $W(M)^*$ dual to $e$, then the canonical Riemannian volume element can be written the form:

$$\omega_a = \mathrm{e}^1 \wedge \ldots \wedge \mathrm{e}^n. \tag{48}$$

A volume-form ω is called *compatible* with the covariant derivative ∇ if [33]

$$\text{div}_\omega \mathbf{w} = (\nabla_\alpha w^\alpha)\omega. \tag{49}$$

The necessary and sufficient conditions to the existence of solutions for eqs.(40) and (49) can be obtained from the following theorem.

**Theorem 1** [23] *Let M be a differentiable manifold endowed with the metric **a** and a covariant derivative $\nabla$. Let $\omega = \varphi \omega_a$, where $\varphi$ is a smooth positive scalar on M, be a volume form on M. Then, for every vector field* $\mathbf{w} = w^\alpha \partial_\alpha \in W(M)$,

$$\nabla_\alpha w^\alpha = \text{div}_\omega \mathbf{w} + \text{t}(\mathbf{w}) + \text{v}(\mathbf{w}) - \mathbf{w}(\ln \varphi), \tag{50}$$

*where* t *and* v *are defined in eq.(46).*

It follows from this theorem that the condition (49) is equivalent to the condition

$$t_\alpha + v_\alpha - \partial_\alpha \ln \varphi = 0. \tag{51}$$

Finally, it follows that, in the Riemannian space $M_a = (M, \mathbf{a})$, a volume (surface if $n = 2$) form $\omega$ is compatible with the covariant derivative $\nabla$ if and only if [23]

$$\omega = e^\varphi \omega_a, \quad t + v = d\varphi. \tag{52}$$

## 4. Geodesics in the material space of graphene sheets

Let $\gamma: I \to M$, $I \subset R_+$ - an interval, be a smooth curve, $\mathbf{v} = v^\alpha \partial_\alpha \in W(M)$, $[\mathbf{v}] = \text{cm}^{-1}$ (or $[\mathbf{v}] = \text{s}^{-1}$), $\tilde{\mathbf{v}} = \mathbf{v} \circ \gamma$ - the rate of change of the curve $\gamma$, and $f \in C^\infty(M)$. Let $(O, u)$, $u = (u^\alpha): O \subset M \to \mathbb{R}^n$, be a coordinate system on M such that $[u^\alpha] = \text{cm}$. Let us denote $P = P(u) \in O$ iff $u = u(P) \in U = u(O)$,

$$\mathbf{v}_{P(u)} \equiv \mathbf{v}(P(u)) = v^\alpha(u) \partial_{\alpha|P} \in T_P M,$$
$$v^\alpha(u) := v^\alpha(P(u)), \quad f(u) := f(P(u)), \tag{53}$$

and

$$\tilde{\mathbf{v}} = \tilde{v}^\alpha \partial_\alpha, \quad \tilde{v}^\alpha = v^\alpha \circ \gamma, \quad \tilde{\mathbf{v}}(t) \in T_{\gamma(t)} M, \quad t \in I,$$
$$\gamma_u = u \circ \gamma = (\gamma^\alpha): I \to \mathbb{R}^n, \quad \gamma^\alpha = u^\alpha \circ \gamma: I \to R, \tag{54}$$

where R denotes the field of real numbers and $\mathbb{R}^n$ denotes the arithmetic vector space over the field R, $[t] = [1]$ (or $[t] = \text{s}$). The rate of change of the function $f$ along the curve $\gamma$ can be written in terms of eqs.(53) and (54) as

$$\frac{df(\gamma(t))}{dt} \equiv \frac{df(\gamma_u(t))}{dt} = \frac{\partial f}{\partial u^\alpha} \frac{d\gamma^\alpha(t)}{dt} = \dot{\gamma}^\alpha(t) \partial_\alpha f,$$
$$\dot{\gamma}^\alpha := \frac{d\gamma^\alpha}{dt}, \quad [\dot{\gamma}^\alpha] = \text{cm}, \quad ([\dot{\gamma}^\alpha] = \text{cms}^{-1}). \tag{55}$$

Particularly

$$\frac{dv^\alpha}{dt} := \frac{d\tilde{v}^\alpha}{dt} = \dot{\gamma}^\beta \partial_\beta v^\alpha. \tag{56}$$

Let us denote

$$\frac{Dv^\alpha}{dt} \equiv \frac{dv^\alpha}{dt} + \Gamma^\alpha_{\beta\kappa} v^\beta \dot{\gamma}^\kappa. \tag{57}$$

The curve $\gamma$ is a $\nabla$-*geodesic* if there exists a field $\mathbf{v} = v^\alpha \partial_\alpha \in W(M)$, $[\mathbf{v}] = \text{cm}^{-1}$ (or $[\mathbf{v}] = \text{s}^{-1}$), $[\partial_\alpha] = \text{cm}^{-1}$, and such positive scalar $h \in C^\infty(M)$ that

$$\frac{\mathrm{D} v^\alpha}{\mathrm{d}t} = 0, \quad \alpha = 1,\ldots,n, \qquad (58)$$
$$v^\alpha = h\dot\gamma^\alpha, \quad [h] = \text{cm}^{-1}, \quad ([h] = \text{cm}^{-1}\text{s}),$$

or, equivalently:

$$\frac{\mathrm{d}^2 \gamma^\alpha}{\mathrm{d}t^2} + \Gamma^\alpha_{\beta\kappa} \frac{\mathrm{d}\gamma^\beta}{\mathrm{d}t} \frac{\mathrm{d}\gamma^\kappa}{\mathrm{d}t} = -\frac{1}{h} \frac{\mathrm{d}h}{\mathrm{d}t} \frac{\mathrm{d}\gamma^\alpha}{\mathrm{d}t}. \qquad (59)$$

where $\left[\Gamma^\kappa_{\alpha\beta}\right] = \text{cm}^{-1}$. A regular differentiable curve $\gamma$ defined by the conditions (57) – (59) is called also a *pregeodesic* [43]. Introducing the new parameter $s$ called a $\nabla$-*geodesic parameter* or *canonical affine parameter* and defined by

$$s = s_0 + c\int \frac{\mathrm{d}t}{h(t)}, \quad [s] = [s_0] = \text{cm}, \quad [c] = [1], \qquad (60)$$

we can rewrite the equation of a geodesic in the following equivalent form:

$$\frac{\mathrm{d}^2 \gamma^\alpha}{\mathrm{d}s^2} + \Gamma^\alpha_{\beta\kappa} \frac{\mathrm{d}\gamma^\beta}{\mathrm{d}s} \frac{\mathrm{d}\gamma^\kappa}{\mathrm{d}s} = 0. \qquad (61)$$

The curve $\gamma$ is called then the *canonically parametrized geodesic* and a parameter $s$ in which the equation of a geodesic takes the above canonical form is called a *natural parameter*. The natural parameter is defined uniquely up to affine transformations:

$$\tau = \tau_0 + cs, \quad [\tau] = [\tau_0] = \text{cm}, \quad [c] = [1]. \qquad (62)$$

Notice that $\nabla$-geodesics depends only on the symmetric part of the $\nabla$-Christoffel symbols, that is, eq.(61) can be reduced to the form

$$\frac{\mathrm{d}^2 \gamma^\alpha}{\mathrm{d}s^2} + \Gamma^\alpha_{(\beta\kappa)} \frac{\mathrm{d}\gamma^\beta}{\mathrm{d}s} \frac{\mathrm{d}\gamma^\kappa}{\mathrm{d}s} = 0. \qquad (63)$$

It follows that the covariant derivatives with the same "symmetric part" $\nabla^s$ defined by

$$\nabla^s_\mathbf{v} \mathbf{u} = \frac{1}{2}(\nabla_\mathbf{v} \mathbf{u} + \nabla_\mathbf{u} \mathbf{v}), \qquad (64)$$

have the same geodesics, and pregeodesics, too [43].

The natural parameter of $\nabla$-geodesics offers a method of defining intervals $s \in \langle 0, l_g \rangle$, $0 < l_g \leq \infty$ along a geodesic which is independent of the metric. If the affine differential manifold $(M, \nabla)$ is additionally endowed with a metric tensor $\mathbf{a}$, then it seems natural, to require that the $\nabla$- geodesic parameter is in agreement with the $\nabla^\mathbf{a}$- geodesic parameter along $\nabla$-geodesics defined by eq.(63). Let $\tau \in I_a \subset \mathbb{R}_+$ be the metric interval along such geodesic and

$$\tau(s) = \int_0^s \|\mathbf{v}(r)\|_a \, \mathrm{d}r, \qquad (65)$$
$$\mathbf{v}(r) = v^\alpha(r)\partial_\alpha, \quad v^\alpha(r) = \dot\gamma^\alpha(r),$$

Then, according to eq.(62), the condition for agreement between the two intervals is (in differential form) [3]:

$$\frac{d^2\tau}{ds^2} = 0. \tag{66}$$

Using the geodesic equation (63), and taking into account eqs.(32) – (36), we obtain that the following condition should be fulfilled along the considered $\nabla$-geodesic [3]:

$$u^\alpha u^\nu u^\sigma W_{\alpha\nu\sigma} = 0, \qquad u^\alpha \equiv \dot{\gamma}^\alpha. \tag{67}$$

For example, according to eq.(36), in the case of the so-called *Weyl-Cartan* geometry defined by the condition

$$\nabla_\kappa a_{\alpha\beta} = w_\kappa a_{\alpha\beta}, \tag{68}$$

eq.(67) reduces to the condition

$$w_\alpha \dot{\gamma}^\alpha = 0, \tag{69}$$

while in the case of the so-called *Riemann-Cartan* geometry:

$$\nabla \mathbf{a} = 0, \tag{70}$$

the condition (67) is fulfilled for any $\nabla$ - geodesic regardless of the choice of metric. In this case, the covariant derivative $\nabla$ is *consistent* with the metric **a** (or – *metrizable*) and, according to eqs.(37) – (39), the nonmetric Christoffel symbols of eq.(32) reduces to the contortion tensor.

**Statement 2** *For each* **a***-metric covariant derivative* $\nabla$ *and for each* $\nabla$*-geodesic we can always choose the* **a***-metric length of this geodesic as its canonical parameter.*

In the case of *isothermal geometry* of corrugated graphene sheets [39] (see also Section 5) we are dealing with the Weyl-Cartan geometry without torsion which can be reduced to the Riemanian geometry for a particular thermal distortions of these sheets leading to the so-called *Weyl integrable geometry* ([39]; see also Section 6) . The Riemannian perfect material geometry of the graphene sheet is admitted in this case (see Section 3). However, it is well known that a crystal with many dislocations reveals the short range order and thus dislocations have no influence on local metric properties of a crystal structure (e.g. [17]; see Section 2). Consequently, for example, the *nonmetricity* can be used as measure of point defects densities (see e.g. [17] and [18]) or as a measure of the existence of another distortions of the graphene crystal structure which produce also a curvature of graphene sheets (see Section 1, remarks at the beginning of Sections 3, and [39] – Section 1).
If the crystal structure is distorted by dislocations only, then the covariant derivative $\nabla$ metric with respect to an internal length measurement tensor **a,** that is defined by eq.(70), is considered (see e.g. [15], [19], [38], [40] - [42]). Then, according to Statement 2, the line elements along $\nabla$-geodesics agree with the line element defined along these geodesics by the metric tensor **a** of the internal length measurement**.** This internal length measurement is, in general, non-Euclidean not only in the case of isothermal geometry of corrugated graphene sheets [39] but also in the case of three-dimensional Bravais crystals. Namely, although in this last case the curvature can be also induced by a macroscopically continuous temperature field (e.g., [18], [28]), nevertheless the influence of curvature-type defects due to the *secondary* *point defects* created by a distribution of many dislocations ought to be taken into account [40] – [42]. Notice that the *curvature-type defects* can occur also in continuized graphene sheets not only as the infinitesimal counterpart of disclinations (see Section 6 – the commentary following eq.(111)) but, according to Conclusion 1 (Section 5), they can appear in the form of topological defects created by a distribution of many edge dislocations in a corrugated graphene sheet. Moreover, curvature-type topological defects can be formed, for example, by replacing a hexagon of the crystalline structure by a *n*-sides polygon (Section 1).

Finally, it ought to be stressed that in the case of corrugated graphene sheets we are dealing, in the contrast to the so far considered continuously defective crystal structures (see e.g. [7]), with the solid body which does not possess an undistorted spatial configuration [39]. Consequently, the problem of consistency of $\nabla$-geodesics and $\nabla^a$-geodesics has the essential meaning.

## 5. Riemann-Cartan material geometry with vectorial torsion

The metric covariant derivative such that for a distinguished vector field $\mathbf{t} \in W(M)$ and for any vector fields $\mathbf{u}, \mathbf{v} \in W(M)$ the following condition is fulfilled:

$$\nabla_{\mathbf{v}} \mathbf{u} = \nabla_{\mathbf{v}}^a \mathbf{u} + (\mathbf{u}, \mathbf{v})_g \mathbf{t} - (\mathbf{t}, \mathbf{u})_g \mathbf{v} \qquad (71)$$

is called a metric covariant derivative with *vectorial torsion*. In this case, according to eq.(159), we have:

$$T(\mathbf{u}, \mathbf{v}) = 2S(\mathbf{u}, \mathbf{v}) = (\mathbf{t}, \mathbf{u})_g \mathbf{v} - (\mathbf{t}, \mathbf{v})_g \mathbf{u} . \qquad (72)$$

The case of two-dimensional manifolds is special because then any metric covariant derivative has vectorial torsion. [1]

The curvature tensor $\mathbf{R}^a$ of the Levi-Civita covariant derivative $\nabla^a$ corresponding to the material metric tensor $\mathbf{a}$ (and treated as a mapping – Appendix) can be written, in the case of a *flat metric covariant derivative* $\nabla$ with vectorial torsion, in the form which explicitly shows how the torsion influences this tensor [2]:

$$R^a(\mathbf{u},\mathbf{v})(\mathbf{w}) = (\mathbf{u},\mathbf{w})_a \nabla_{\mathbf{v}}\mathbf{t} - (\mathbf{v},\mathbf{w})_a \nabla_{\mathbf{u}}\mathbf{t} + (\nabla_{\mathbf{u}}\mathbf{t} + \|\mathbf{t}\|_a^2 \mathbf{u}, \mathbf{w})_a \mathbf{v} - (\nabla_{\mathbf{v}}\mathbf{t} + \|\mathbf{t}\|_a^2 \mathbf{v}, \mathbf{w})_a \mathbf{u} , \qquad (73)$$

where eq.(162) was taken into account. Because, in is the case, the curvature tensor of the Riemannian material space $M_a = (M, \mathbf{a})$ vanishes if the torsion vanishes, it leads to the following conclusion:

**Conclusion 1** *The case of flat metric covariant derivative $\nabla$ in the material space M of a corrugated graphene sheet is physically admissible if*
(i) *The Riemannian material space $M_a = (M, \mathbf{a})$ is flat and dislocations are absent. In this case the graphene sheet can be observed in the physical configurational space as a developable surface.*
(ii) *Edge dislocations are the only source of non-flatness of the Riemannian material space. Consequently, it is the case of secondary curvature-type defects created by the distribution of these dislocations.*

The case of three-dimensional Bravais crystals suggests the following hypothesis (see remarks at the end of Section 4):

**Hypothesis 2** *The secondary curvature-type defects created by a distribution of many edge dislocations in a corrugated graphene sheets can appear in crossover points of edge dislocation lines or when two parallel dislocation lines are joined together.*

The coordinate description of the vectorial torsion has the form
$$S_{\alpha\beta}{}^{\kappa} = t_{[\alpha}\delta^{\kappa}_{\beta]},$$
$$T_{\alpha\beta}{}^{\kappa} = 2S_{\alpha\beta}{}^{\kappa}, \quad t_{\alpha} = a_{\alpha\beta}t^{\beta} = T_{\alpha\kappa}{}^{\kappa}, \tag{74}$$

where eqs.(34) and (46) were taken into account. Moreover, it follows from eqs.(19) and (74) that would be:
$$\mathbf{t} = t^{\alpha}\partial_{\alpha}, \quad [\mathbf{t}] = \text{cm}^{-2}, \quad [t^{\alpha}] = [\partial_{\alpha}] = \text{cm}^{-1}. \tag{75}$$

The components of the curvature tensor $\mathbf{R}$ have the form (Appendix):
$$R_{\kappa\alpha\beta}{}^{\nu} = \partial_{\kappa}\Gamma^{\nu}_{\alpha\beta} - \partial_{\alpha}\Gamma^{\nu}_{\kappa\beta} - \left(\Gamma^{\rho}_{\kappa\beta}\Gamma^{\nu}_{\alpha\rho} - \Gamma^{\rho}_{\alpha\beta}\Gamma^{\nu}_{\kappa\rho}\right). \tag{76}$$

The components of the Ricci tensor $\mathbf{R}_c$ of Statement 1 are defined as
$$R_{\alpha\beta} := R_{\kappa\alpha\beta}{}^{\kappa} = \partial_{\kappa}\Gamma^{\kappa}_{\alpha\beta} - \partial_{\alpha}\Gamma^{\kappa}_{\kappa\beta} - \left(\Gamma^{\rho}_{\kappa\beta}\Gamma^{\kappa}_{\alpha\rho} - \Gamma^{\rho}_{\alpha\beta}\Gamma^{\kappa}_{\kappa\rho}\right), \tag{77}$$

and the following relation holds [43]:
$$R_{\alpha\beta} - R_{\beta\alpha} = -R_{\alpha\beta\kappa}{}^{\kappa} = \sum_{\kappa=1}^{n}\left(\partial_{\alpha}\Gamma^{\kappa}_{\kappa\beta} - \partial_{\beta}\Gamma^{\kappa}_{\kappa\alpha}\right). \tag{78}$$

In the presence of the curvature and torsion, which describe indirectly the macroscopically small translational and rotational discrepancies, the second covariant derivatives do not commute and satisfy the *Ricci identities*:
$$\left(\nabla_{\beta}\nabla_{\alpha} - \nabla_{\alpha}\nabla_{\beta}\right)f = T_{\alpha\beta}{}^{\lambda}\nabla_{\lambda}f,$$
$$\left(\nabla_{\beta}\nabla_{\alpha} - \nabla_{\alpha}\nabla_{\beta}\right)u^{\kappa} = R_{\beta\alpha\lambda}{}^{\kappa}u^{\lambda} - T_{\alpha\beta}{}^{\lambda}\nabla_{\lambda}u^{\kappa}, \tag{79}$$

where $\mathbf{u} = u^{\alpha}\partial_{\alpha} \in W(M)$, $f \in C^{\infty}(M)$.

"An interesting fact is that, in a Riemann-Cartan space, there are transformations involving both the metric and the covariant derivative, which preserve the metric compatibility condition. In addition, these transformations leave invariant the curvature and, at the same time, change the torsion in a way similar to a gauge transformation. They are defined by the combined effect of a *conformal transformation* of the metric
$$\tilde{a}_{\alpha\beta} = e^{\vartheta}a_{\alpha\beta}, \quad \vartheta = -2\sigma, \tag{80}$$

where $\sigma \in C^{\infty}(M)$, $[\sigma] = [1]$, and by a particular case of transformations that preserve all $\nabla$-geodesics called *Einstein's $\lambda$-transformations* of the Christoffel coefficients of the metric covariant derivative $\nabla$:
$$\tilde{\Gamma}^{\kappa}_{\alpha\beta} = \Gamma^{\kappa}_{\alpha\beta} + \frac{1}{2}\partial_{\alpha}\vartheta\delta^{\kappa}_{\beta}, \quad \vartheta = -2\sigma. \tag{81}$$

Note that the above transformations do not involve coordinate transformations. It easy to verify that under these transformations, the components of the torsion transform according to:

$$\tilde{T}_{\alpha\beta}{}^{\kappa} = T_{\alpha\beta}{}^{\kappa} - 2\partial_{[\alpha}\sigma\delta_{\beta]}^{\kappa}. \tag{82}$$

Conversely, under both the conformal transformation of the metric (80) and the transformation of torsion (82) the ∇-Christoffel coefficients transform according to an Einstein's λ-transformation (81)."[8]

It follows from eq.(74) that the transformations (80) and (82) preserve also the vectorial type of torsion and thus these are *canonical Riemann-Cartan transformations* [8] which preserve, in the case of continuously defective graphene sheets, the kind of their material geometry. The components of curvature and Ricci tensors are left invariant under this transformation [8]:

$$\tilde{R}_{\kappa\alpha\beta}{}^{\nu} = R_{\kappa\alpha\beta}{}^{\nu}, \qquad \tilde{R}_{\alpha\beta} = R_{\alpha\beta} \tag{83}$$

but the curvature scalar

$$R = a^{\alpha\beta}R_{\alpha\beta} \tag{84}$$

transforms as [8]:

$$\tilde{R} = e^{2\sigma}R. \tag{85}$$

As it is known, the torsion tensor can be decomposed into two parts:

$$T_{\alpha\beta}{}^{\kappa} = L_{\alpha\beta}{}^{\kappa} + \frac{1}{n-1}\left(t_{\alpha}\delta_{\beta}^{\kappa} - t_{\beta}\delta_{\alpha}^{\kappa}\right),$$

$$L_{\alpha\kappa}{}^{\kappa} = 0, \qquad t_{\alpha} = T_{\alpha\kappa}{}^{\kappa}. \tag{86}$$

It follows from eqs.(81), (82) and (86) that this decomposition transforms, under these canonical transformations, as

$$\tilde{T}_{\alpha\beta}{}^{\kappa} = \tilde{L}_{\alpha\beta}{}^{\kappa} + 2\tilde{t}_{[\alpha}\delta_{\beta]}^{\kappa},$$

$$\tilde{L}_{\alpha\beta}{}^{\kappa} = L_{\alpha\beta}{}^{\kappa}, \qquad \tilde{t}_{\alpha} = t_{\alpha} - (n-1)\partial_{\alpha}\sigma. \tag{87}$$

We see that the trace of the torsion plays the role of a gauge vector field, whereas the traceless part is invariant.[8]

If $n = 2$, then the torsion is vectorial and, comparing eqs.(74) and (86), we obtain that, in the case of the Riemann-Cartan material space of a corrugated graphene sheet, the canonical Riemann-Cartan transformations lead to the following relations:

$$L_{\alpha\beta}{}^{\kappa} = 0, \qquad \tilde{t}_{\alpha} = t_{\alpha} - \partial_{\alpha}\sigma. \tag{88}$$

It follows from eqs.(35) – (38), (49) – (52), (70) and (86) – (88) that the area form ω is *compatible* with the metric covariant derivative ∇ (Section 3) if and only if the torsion is the so-called *gradient vectorial torsion* (cf. eq.(52)):

$$\mathbf{t} = \mathrm{d}\varphi, \tag{89}$$

or, equivalently (see [39] – Appendix C),

$$\mathbf{t} = \mathrm{grad}_a \varphi. \tag{90}$$

This compatible area form is given, according to eqs. (15), (27), (28), and (52), by the following formula:

$$\omega = e^{\varphi}\omega_a = e^{\varphi}\sqrt{a}\,\mathrm{d}u^1 \wedge \mathrm{d}u^2, \tag{91}$$

and it is the *area element* iff the condition (16) is fulfilled (see Appendix – (152)). The torsion tensor is defined by eqs.(74) and (89) and it follows from eqs.(32), (37)-(39), (74), and (89) that the difference tensor has the form

$$Q_{\alpha\nu\sigma} = K_{\alpha\nu\sigma} = 2\partial_{[\alpha}\varphi a_{\nu]\sigma}. \tag{92}$$

Finally, the Riemann-Cartan material space *M* (of the corrugated graphene sheet) possessing a gradient torsion is characterized by two fundamental objects, namely the metric **a** and the scalar field $\varphi$ which transform according to the following *compatibility transformations* [8]:

$$\tilde{\mathbf{a}} = e^{-2\sigma}\mathbf{a}, \qquad \tilde{\varphi} = \varphi - \sigma. \tag{93}$$

In the case of a Riemann-Cartan space with vectorial torsion, the following theorem holds:

**Theorem 2** ([1], [2]) *Let $M_a = (M, \mathbf{a})$ be a 2-dimensional Riemannian manifold with Gaussian curvature K.*
(i) *The **a**-metric covariant derivative $\nabla$ with vectorial torsion is flat if and only if*

$$K = \text{div}_a \mathbf{t} \tag{94}$$

*holds. In particular, if M is compact, then M is diffeomorphic to torus or Klein bottle.*
(ii) *If M is connected, complete and simply connected and **t** is $\nabla$-parallel, i.e.,*

$$\nabla \mathbf{t} = 0, \tag{95}$$

*then M is non-compact space of constant negative Gaussian curvature given by*

$$K = -\|\mathbf{t}\|_a^2, \tag{96}$$

*that is, $M_a$ has to be isometric to hyperbolic space.*
(iii) *Let $\varphi$ be a smooth function on the Riemannian manifold $M_a = (M, \mathbf{a})$, $\nabla$ - the covariant derivative with gradient vectorial torsion (eq.(90)), and consider the conformally equivalent metric $\tilde{\mathbf{a}}$ defined by eq.(93). Then any $\nabla$ - geodesic $\gamma(t)$ is, up to a reparametrization $\tau = \tau(t)$, a $\nabla^{\tilde{a}}$ - geodesic, and the function $\tau$ is the unique solution of the differential equation*

$$\ddot{\tau} - \dot{\tau}\dot{\sigma} = 0,$$
$$\sigma(t) := \sigma \circ \gamma \circ \tau(t). \tag{97}$$

It follows from the above theorem that:

**Statement 3** *In the case of Riemann-Cartan material geometry with the gradient vectorial torsion (eq.(90)):*
(i) *$\nabla$-geodesics cover, up to the reparametrization (97), with $\tilde{\mathbf{a}}$ - geodesics defined by the compatibility transformation (93).*
(ii) *This Riemann-Cartan material space is flat iff the Gaussian curvature K of the Riemannian material space $M_a = (M, \mathbf{a})$ and the potential $\varphi$ governing the distribution of dislocations are interrelated according to the following equation:*

$$\Delta_a \varphi = K, \tag{98}$$

*where $\Delta_a$ denotes the Laplacian operator corresponding to the metric **a** ( [39] - Appendix C).*

The *flatness* of a Riemann-Cartan space means that the covariant derivative $\nabla$ covers (at least locally) with the covariant derivative $\nabla^e$ of a *teleparallelism* on $M$ defined by a base $e = (\mathbf{e}_a; a = 1,\ldots,n)$ (n = 2 in our case) of the module W($M$) of smooth vector fields on $M$ tangent to $M$ (Appendix). This covariant derivative is called also the *Weitzenbök* covariant derivative and is uniquely defined by the conditions (153) and (165). The components $S_{ab}{}^c$ of the torsion tensor of this covariant derivative are defined by the relation

$$\tau^a = de^a = S_{bc}{}^a e^b \wedge e^c, \qquad S_{bc}{}^a = \frac{1}{2} T^a{}_{bc}, \tag{99}$$

where, according to eqs.(153), (156), (160), and (165), we have:

$$T_{ab}{}^c = -C_{ab}^c. \tag{100}$$

It follows from eqs.(72), (100), (160), (165) that in the case of *vectorial torsion* we have

$$C_{ab}^c = t_b \delta_a^c - t_a \delta_b^c. \tag{101}$$

We will say that the vector base $e = (\mathbf{e}_a; a = 1,\ldots,n)$ fulfilling the condition (101) describes a *vectorial teleparallelism* on $M$. Particularly, if $n = 2$, then eq.(30) takes the form

$$\rho \mathbf{b} = [\mathbf{e}_1, \mathbf{e}_2] = t_2 \mathbf{e}_1 - t_1 \mathbf{e}_2. \tag{102}$$

Notice that eq.(31), the above orientation of the local Burgers vector and the orientation of the manifold $M_a$, are invariant with respect to the local rotations, $e = (\mathbf{e}_a) \to \tilde{e} = (\tilde{\mathbf{e}}_a)$, where

$$\begin{aligned} \tilde{\mathbf{e}}_1 &= \cos\vartheta \mathbf{e}_1 + \sin\vartheta \mathbf{e}_2, \\ \tilde{\mathbf{e}}_2 &= -\sin\vartheta \mathbf{e}_1 + \cos\vartheta \mathbf{e}_2. \end{aligned} \tag{103}$$

Hence

$$\omega = e^1 \wedge e^2 = \tilde{e}^1 \wedge \tilde{e}^2. \tag{104}$$

However, the transformation (103) in general does not preserve the relation (102). For example, if $t_a = \text{const.} \neq 0$, $a = 1, 2$, then a constant rotation transforms the commutator into the form

$$[\tilde{\mathbf{e}}_1, \tilde{\mathbf{e}}_2] = t_2 \tilde{\mathbf{e}}_1. \tag{105}$$

It can be shown, that if the condition (41) of **a**-orthonormality of the base $e = (\mathbf{e}_a; a = 1,\ldots,n)$ is fulfilled and the Lie bracket of eq.(156) is defined by the object of anholonomity $C_{ab}^c$ of the form (101), then, basing oneself on the formulas taken from [4] and [40], we obtain the following *self-balance equation*:

$$\text{div}_a \mathbf{e}_b \doteq t_b, \qquad t_b = C^a{}_{ab} (\equiv C_{ab}^a), \tag{106}$$

where div$_a$ is the divergence operator defined on the Riemannian space $M_a$ and $\doteq$ means that the form of the relation (106) needs an **a**-orthonormal base *e*. Notice that a continuized graphene sheet with the ideal material geometry is an isotropic continuous planar solid body (see remarks at the beginning of Section 3). On the other hand, in the case of the flat Riemann-Cartan geometry, we are dealing with the *secondary curvature-type defects* of graphene sheets only (Conclusion 1 and Statement 3). Thus, from the point of view of the description of continuously dislocated graphene sheets, it seems physically sensible to distinguish the case of flat Riemann-Cartan geometry.

Let us consider, in order to compare this material geometry with the *isothermal geometry of corrugated graphene sheets* [39], a two-dimensional differential manifold *M* endowed with a pair (**a**, w) which defines the Weyl geometry of *M* (see eqs.(36), (68) and e.g. [32]), that is, **a** is a Riemannian metric tensor on *M* and $w = w_\alpha du^\alpha \in W(M)^*$ is a distinguished 1-form on *M*. It is assumed that *M* is additionally endowed with a smooth field $\theta \in C^\infty(M)$ of effective absolute temperatures. It can be shown [39] that there are *variational material field equations* admitting to formulate the following *isothermal thermal state equation*:

$$\mathrm{div}_a \mathbf{w} + 2K = r, \qquad r = const.,$$
$$\mathbf{w} = w^\alpha(u, \theta, d\theta)\partial_\alpha, \quad w^\alpha = a^{\alpha\beta}w_\beta, \quad w_\beta = 2\varepsilon_\beta, \tag{107}$$

where *K* is the Gauss curvature of the material Riemannian space $M_a = (M, \mathbf{a})$ and *r* is any constant belonging to a certain countable set of real numbers. This equation states that the corrugations of graphene sheets are interrelated with their thermal state and in this sense we are dealing with the isothermal geometry of these sheets. In this case the antisymmetric tensor field

$$F_{\alpha\beta} = \partial_\alpha \varepsilon_\beta - \partial_\beta \varepsilon_\alpha, \tag{108}$$

analogous to the effective electromagnetic field, plays the role of a measure of the influence of the graphene effective temperature on the length measurement in the Weyl material space of corrugated graphene sheets.[39] We can now formulate the following conclusion describing in details the considered particular case of material geometry.

**Conclusion 2** *If the Riemannian material space $M_a = (M, \mathbf{a})$ of a graphene sheet is given, and the **a**-metric covariant derivative $\nabla$ is flat, then there exists a vector field $\mathbf{t} \in W(M)$ which defines the Gaussian curvature of the space $M_a$ (eq.(94)) as well as the vectorial torsion of $\nabla$ (eq.(72)). In this particular case we have*:

(i) *The covariant derivative $\nabla$ covers (at least locally) with the covariant derivative $\nabla^e$ of a teleparallelism on M defined by a base e = ($\mathbf{e}_a$; a = 1, 2) and thus this moving frame defines a vectorial teleparallelism. Consequently, eqs.(100) – (102) hold and the local Burgers vector **b**, the scalar density of dislocations ρ (Definition 1 and eq.(102)) as well as the corresponding congruence of virtual effective edge dislocation lines (Definition 2) are defined by this metric covariant derivative and the moving frame e. Moreover, if the moving frame e is **a**-orthonormal, then the self-balance equation (106) holds.*

(ii) *The Gaussian curvature K of the Riemannian material space $M_a$ can be interpreted, according to eq.(94), as the "charge density" of vectorial torsion. It follows that this curvature equals zero when the effective edge dislocations are absent. Conversely, if the Gaussian curvature vanishes, that is the secondary curvature-type defects are absent, then we are dealing with a continuous distribution of edge dislocations located in a developable graphene sheet. If additionally the torsion is a gradient vectorial torsion, then the condition K = 0 defines a harmonic potential φ (eq.(98)) governing this distribution of dislocations.*

(iii) *The comparison of the material Riemannian space $M_a$ with the integrable isothermal geometry of corrugated graphene sheets, represented by eq.(107) and by the condition $F_{\alpha\beta} = 0$ in eq.(108) (see also remarks in Section 4), leads to the conclusion that if the graphene sheet* *is not compact, then eq.(94) can be taken as the compatibility condition of these material geometries.*

We will see in the next section that if the Riemann-Cartan material geometry is nonplanar, then the existence of torsion-type defects in a corrugated graphene sheet with this material geometry can be closely related with the existence of a characteristic length parameter of the material structure of this defective graphene sheet.

## 6. Variational material geometry and related topics

"Elastic behaviour of matter is usually classified as reversible, inelastic behaviour (associated with the existence of plastic deformations) – as irreversible. Like in thermodynamics, irreversibility in mechanics is much more involved than reversibility….It is a distinctive feature of inelasticity that the decisive motion processes occur in the *interior of bodies*. These processes are either elementary or composed: they often involve *small scale defects* in the material constitution. The defects, too, are classified as elementary or composed."([20]; see also Section 1, Conclusion 1, and Hypothesis 2). This quotation explain well the reason why we assume that the *material space* of graphene sheets, understood as a surface endowed with the material geometry independent from the geometry of the configurational space, is virtually the glide surface along which *effective edge dislocations* can move (Sections 2 and 3; see also [39] and [40] – [42] in the case of three-dimensional crystals). So, some aspects of irreversible behaviour of continuized crystals can be described in terms of the material space. It is, in the naming of the paper [19], the perception of the *internal observer* of the state of a crystal. It differs from the perception of "the *external observer* who looks into the crystal from outside, namely from the Euclidean space in which the crystal is embedded."[19] We remind that external observers of two kinds exist in the case of corrugated graphene sheets: the observer of mechanical phenomena located in the 3D Euclidean point space and the observer of electronic phenomena located in the (2D +1) Minkowski point space-time.[39]

Let us consider the following "Gedanken Experiment" (see e.g. [19] and [38]). Imagine that an ideal graphene sheet embedded in the Euclidean space is cut into infinitesimal surface elements each of which is then subject to a stress-free distortion (i.e., strain plus rotation) that varies from one element to the other in such a way that the distorted lattice is uniquely defined everywhere (i.e., it is locally a perfect lattice) and that the misfit arises in a *macroscopically continuous* manner (see the beginning of Section 2).

The question appears: what really happens when the stress-free distortions are applied to the original ideal crystal? Because the *ideal graphene* is the isotropic material and we are dealing with a *locally perfect* graphene sheet, the *translational discrepancy* appearing in the above Gedanken Experiment can be described with the help of a distinguished *moving frame* $e = (e_a; a = 1,2)$ being a base of the module W(*M*) (eq. (27)) and a metric tensor **a** of eq.(17) such that the condition (41) is fulfilled (i.e., the moving frame $e$ is defined up to *local rotations* of eq.(103)). Next, the *rotational discrepancy* appearing in this Gedanken Experiment can be represented by infinitesimal relative rotations of the base vector fields $e_a$. These relative rota-tions can be described with the help of a covariant derivative ∇ assigning to the moving frame $e$ a matrix function of 1-forms ($\omega^a{}_b$) with values in the Lie algebra so(2) of infinitesimal rotations. (cf. [38] where the isotropic and tranversally isotropic continuously dislocated three-dimensional crystals are considered).

The physical reason to consider the above defined triple ($e$, **a**, ∇) of geometrical objects is that the family of distorted *macroscopically infinitesimal elements* (see Section 2) of an ideal graphene sheet mismatched in Euclidean geometry, fit perfectly in the geometry defined by these objects. Namely, let $M_a = (M, \mathbf{a})$ be a material Riemannian space representing the *internal length measurement* in the corrugated graphene sheet (eqs.(17), (18), (20)) and let ∇ be the Riemann-Cartan covariant derivative consistent in this internal length measurement in the sense of the condition (70). We have then the following *compatibility relation* of the considered geometrical objects:

$$\nabla \mathbf{e}_a = \omega^b{}_a \otimes \mathbf{e}_b = \omega \otimes \gamma_a \in W(M)^* \otimes W(M),$$
$$\omega^a{}_b = \varepsilon^a{}_b \omega = \omega_\kappa{}^a{}_b du^\kappa, \quad \omega = \omega_\kappa du^\kappa, \quad \omega_\kappa{}^a{}_b = \omega_\kappa \varepsilon^a{}_b, \quad (109)$$
$$\gamma_a = \varepsilon^b{}_a \mathbf{e}_b, \quad \varepsilon^a{}_b = \delta^{ac} \varepsilon_{cb}, \quad \varepsilon_{cb} = -\varepsilon_{bc}, \quad (\varepsilon^a{}_b) \in so(2),$$

where the condition (41) was taken into account, so(2) is the Lie algebra of the proper rotation group SO(2) and $\varepsilon^{ab}$ denotes the permutation symbol, that is, i.e., $\varepsilon^{12} = -\varepsilon^{21} = 1$, $\varepsilon^{11} = \varepsilon^{22} = 0$. It follows that the antisymmetry condition

$$\omega^a{}_b = -\omega^b{}_a \quad (110)$$

is equivalent to the metricity condition (70). The components $\omega_\kappa{}^a{}_b$ of the 1-forms $\omega^a{}_b$ are called the *Ricci coefficients* of infinitesimal relative rotations (of local directions defined by the base vector fields $\mathbf{e}_a$).

Let's remind, that the existence of effective edge dislocations can be modelled by means of the base $e = (\mathbf{e}_a)$ and the covariant derivative ∇ such the conditions (70) – (72), (99) - (101), (156), and (157) are fulfilled. The virtual effective edge dislocations constitute then the congruence $C[\mathbf{l}]$ of lines defined by the condition (31) (see [40] – [42] in the case of three-dimensional continuously dislocated Bravais crystals). Next, let us consider the particular case when the base $e = (\mathbf{e}_a)$ of W($M$) defines a *closed teleparellelism*, that is, the condition (166) is additionally fulfilled. It is equivalent to the condition that this moving frame spans the $n$-dimensional real Lie algebra $g$ of $e$-parallel vector fields defined by eq.(167). The corresponding distribution of dislocations is called *uniformly dense* and, in the case $n = 3$, can be treated as *massless fundamental states* of the distorted Bravais structure defined by field equations which resemble the equations electrodynamics [37]. Only two types of Lie algebras are possible in two dimensions: the Lie algebra $g$ isomorphic with the Lie algebra so(2) of two-dimensional rotations, or the case when the base vector fields commute:

$$[\mathbf{e}_1, \mathbf{e}_2] = 0 \quad (111)$$

what means, according to eq.(102), that dislocations are absent. When both structure constants of eq.(102) are non-zero, a constant rotation transforms the commutator into the form (105). Further, there are no Jacobi conditions (eq.(158) with $C^c_{ab} = $ constant ) in two dimensions; any two constants can be structure constants.[35] Next, the system of lines, defined by the base vector fields $\mathbf{e}_1$ and $\mathbf{e}_2$ of the Lie algebra isomorphic with so(2), possesses rotational symmetries only and thereby can be considered as the uniformly dense distribution of dislocations of rotational type being a continual counterpart of discrete disclinations (see the paper [31] concerning the topological classification of discrete line defects). Therefore, according to this definition, disclinations are, in the continuous limit, rather a type of continuous distribution of dislocations than a separate kind of line defects.[38] Consequently, in contrast to the discrete case (see Section 1 and [39] – Section 1), we can only say that the continuously defective corrugated graphene sheets reveal *curvature-type defects* which can appear, as it was mentioned in Sections 1 - 3, due to the different physical reasons, and *torsion-type defects* represented by the edge-type effective line defects (Section 3); see also e.g. remarks in [18] and [20] concerning this topic in the case of three dimensional crystals.

The metric coefficients $a_{\alpha\beta}$ of eq.(17) and the Christoffel symbols $\Gamma^{\kappa}_{\alpha\beta}$ of the **a**-metric covariant derivative $\nabla$ of eq.(109) (see eqs.(6), (70), and Section 5) are expressed in terms of the moving frame $e = (\mathbf{e}_a; a = 1, 2)$ of eq.(27) ( a *"zweibein"* as it is called frequently in the literature) by

$$\nabla_\beta e^a{}_\alpha = \partial_\beta e^a{}_\alpha - \Gamma^{\kappa}_{\beta\alpha} e^a{}_\kappa - \omega_\beta{}^{ab} e_{b\alpha} = 0,$$
$$a_{\alpha\beta} = e^a{}_\alpha e^b{}_\beta \delta_{ab}, \quad \omega_\alpha{}^{ab} = \omega_\alpha e^{ab}, \quad e_{b\alpha} = a_{\alpha\beta} e_b{}^\beta, \qquad (112)$$

where $\varepsilon^{ab}$ ($= \varepsilon_{ab}$) is the permutation symbol associated with the **a**-orthonormal moving frame $e$. Let us denote by $e_{\alpha\beta}$ and $e^{\alpha\beta}$ the covariant 2-vector density of weight -1 and the contravariant 2-vector density of weight +1, respectively, defined by:

$$e_{\alpha\beta} = \sqrt{a}\varepsilon_{\alpha\beta}, \quad e^{\alpha\beta} = \frac{1}{\sqrt{a}}\varepsilon^{\alpha\beta}, \quad a = \det(a_{\alpha\beta}), \qquad (113)$$

where $\varepsilon^{\alpha\beta} = \varepsilon_{\alpha\beta}$ is the permutation symbol associated with the coordinate system $u = (u^\alpha)$, $[u^\alpha]$ = cm. Notice that

$$e^{\alpha\beta} = e_a{}^\alpha e_b{}^\beta e^{ab}, \quad e_{\alpha\beta} = e^a{}_\alpha e^b{}_\beta e_{ab},$$
$$e^{ab} \doteq \varepsilon^{ab}, \quad e_{ab} \doteq \varepsilon_{ab}, \qquad (114)$$

where eqs.(17), (27), (41), (43), and (45) were taken into account, and [46]

$$e^{\alpha\sigma} e_{\beta\sigma} = e^{\sigma\alpha} e_{\sigma\beta} = \delta^\alpha_\beta,$$
$$e^{\alpha\beta} e_{\lambda\sigma} = \delta^{\alpha\beta}_{\lambda\sigma} \equiv \delta^\alpha_\lambda \delta^\beta_\sigma - \delta^\beta_\lambda \delta^\alpha_\sigma, \qquad (115)$$
$$e_{\alpha\beta} e_{\lambda\sigma} = a_{\alpha\lambda} a_{\beta\sigma} - a_{\alpha\sigma} a_{\beta\lambda}.$$

Next [14], in terms of local coordinates $u = (u^\alpha; \alpha = 1, 2)$, $[u^\alpha]$ = cm, the torsion tensor is given by eq.(74), the curvature tensor of the **a**-metric connection $\nabla$ (eq.(76)) is given by the following expression (in designations of eq.(164); cf. [14] and [47]):

$$R_{\alpha\beta\mu\sigma} = a_{\alpha\kappa} R^{\kappa}{}_{\beta\mu\sigma} = \frac{R}{2} e_{\alpha\beta} e_{\mu\sigma}, \quad R = a^{\alpha\beta} R_{\alpha\beta}, \qquad (116)$$

where the scalar curvature $R \in C^\infty(M)$ and the covector $\mathrm{t} = t_\alpha \mathrm{d} u^\alpha$ of eq.(74) are defined as follows [14]:

$$R := -2 e^{\alpha\beta} \partial_\alpha \omega_\beta, \quad t_\alpha := 2 e_c{}^\kappa \tau^c{}_{\alpha\kappa} - \omega^\kappa e_{\alpha\kappa}, \quad \omega_\alpha = a_{\alpha\kappa} \omega^\kappa, \qquad (117)$$

Let us consider the following action:

$$\mathbb{S} = \int_M L(R,T) \, dF, \quad (118)$$

where eqs.(27), (28) and (45) were taken into account, and the Lagrangian $L$ is assumed in the form [14]:

$$L(R,T) = \frac{1}{4}\left(\sigma R^2 + 2\mu T^2\right) + \lambda, \quad T = \|\mathbf{t}\|_a = \sqrt{t_\alpha t^\alpha}, \quad (119)$$

where $\sigma$, $\mu$ and $\lambda$ are arbitrary constants:

$$\sigma > 0, \quad \mu > 0, \quad \lambda \geq 0. \quad (120)$$

The action $\mathbb{S}$ is positive definite for the Riemannian metrics and it is "the most general action yielding second second-order Euler-Lagrange equations for zweibein and SO(2)-connection which are considered as independent variables. For a fixed $M$, finite and positive definite action one has a well-defined variational problem…Varying the action (118) with respect to SO(2)-connection and zweibein one obtains the Euler-Lagrange equations:" [14]

$$\begin{aligned} &\sigma \nabla_\alpha R + \mu t_a = 0, \\ &\mu \nabla_\alpha t_\beta + \frac{1}{4} a_{\alpha\beta}\left(\sigma R^2 + 2\mu T^2 - 4\lambda\right) = 0, \end{aligned} \quad (121)$$

where $\nabla$ denotes the Riemann-Cartan covariant derivative of eqs.(109) and (112). It follows from eqs.(74), (75) and (119) that if dislocations are absent (torsion tensor vanishes), then the scalar curvature $R$ is constant and the Lagrangian has the form $L(R) = \sigma R^2/4 + \lambda$ being a particular case of the Lagrangian describing the isothermal geometry of corrugated graphene sheets [39]. The value of the square of the constant curvature $R_0$ determined by the above field equations:

$$R_0^2 = \frac{4\lambda}{\sigma} \quad (122)$$

agrees in this particular case with the value of this scalar determined by the field equations describing the isothermal geometry. So, the above field equations are consistent with the field equations of isothermal geometry proposed in [39].

To simplify the form of the equations we introduce, following [14], *dimensionless coordinates*. Namely, assuming that

$$[u^\alpha] = [du^\alpha] = \text{cm}, \quad [\mathbf{e}_a] = [\partial_\alpha] = \text{cm}^{-1}, \quad [a_{\alpha\beta}] = [1], \quad (123)$$

we obtain that

$$\begin{aligned} &[\mathbf{a}] = \text{cm}^2, \quad [R] = \text{cm}^{-2}, \quad [T] = \text{cm}^{-1}, \\ &[\omega_\alpha] = [\omega^\alpha] = \text{cm}^{-1}, \quad [\Gamma^\kappa_{\alpha\beta}] = [\omega_\alpha{}^{ab}] = \text{cm}^{-1}, \end{aligned} \quad (124)$$

and

$$[L] = [\lambda] = [\mu]\text{cm}^{-2} = [\sigma]\text{cm}^{-4}, \quad [L] = [\mathbb{S}]\text{cm}^{-2}, \quad (125)$$

where the action $\mathbb{S}$ has the dimension of energy (i.e., kgcm$^2$s$^{-2}$) multiplied by the time dimension (i.e., s). Thus, under the conditions (120), (123) and (125), dimensionless curvilinear coordinates

$$x = \frac{u^1}{l_0}, \quad y = \frac{u^2}{l_0}, \quad l_0 = \sqrt{\frac{2\sigma}{\mu}}, \quad [l_0] = \text{cm}, \quad (126)$$

and dimensionless complex coordinates

$$z = x + iy, \quad \bar{z} = x - iy, \tag{127}$$

can be defined. These dimensionless coordinates and the coupling constants enter the Euler-Lagrange equations only through one dimensionless constant [14]:

$$\Lambda = \frac{4\lambda\sigma}{\mu^2}. \tag{128}$$

**Statement 4** *If the tensor $F_{\alpha\beta}$ of eq.(108) vanishes, then the Weyl material isothermal geometry of corrugated graphene sheets ([39] and Conclusion 2 – (iii)) reduces to the temperature dependent conformal rescaling ([39] – Section 4) and the length parameter $l_0$ can be identified with the characteristic thermal length parameter $l(\theta_0)$, where $\theta_0 > 0$ is a reference effective absolute temperature [39].*

Notice that if dislocations are absent, then the term $\mu T^2/2$ in the Lagrangian $L$ is absent and thus the characteristic constants $l_0$ and $\Lambda$ cannot be introduced. However, if the influence of dislocations on the material geometry is taken into account, then we can consider $T = 0$ as a particular case and thus these characteristic parameters can be introduced. It follows from eqs.(121), (122) and (128) that then the following relationship holds:

$$R_0^2 = \left(\frac{\mu}{\sigma}\right)^2 \Lambda. \tag{129}$$

It is well known (e.g. [32]) that each two-dimensional Riemannian manifold $M_a = (M, \mathbf{a})$ is *conformally flat,* that is., there exists a local coordinate system $u = (u^1, u^2): O \to \mathbb{R}^2$, $[u^\alpha] =$ cm, $O \subset M$, where $\mathbb{R}^2$ denotes the arithmetic real space $\mathbb{R}^2$ considered as the Euclidean vector space, such that in eq.(17):

$$a_{\alpha\beta} \doteq e^{-2\varphi}\delta_{\alpha\beta}, \tag{130}$$

where $\varphi \in C^\infty(O)$ is a scalar. Further on, for the simplicity of the notation, the images $u(P) = (u^1(P), u^2(P)) \in U \subset \mathbb{R}^2$, $U = u(O)$, of points $P \in O$ under the mapping $u$, are designated also by $u = (u^1, u^2)$. The coordinate system $u = (u^\alpha)$ of eq.(130) can then be identified with a Cartesian coordinate system on the plane $\mathbb{R}^2$. If the metric tensor $\mathbf{a}$ is defined by eqs.(17) and (130), then the general $\mathbf{a}$-orthonormal moving frame $e = (\mathbf{e}_a\, ; a = 1, 2)$ preserving the orientation of $M_a$ can be written in the form:

$$\mathbf{e}_a \doteq e^\varphi \delta_a^\kappa \mathbf{a}_\kappa, \tag{131}$$

where the base vectors $\mathbf{a}_\alpha : U \to W(O)$, $\alpha = 1, 2$, have the form:

$$\mathbf{a}_\alpha = Q_\alpha{}^\beta \partial_\beta, \quad \partial_\beta = \partial/\partial u^\beta,$$
$$\mathbf{Q} = \|Q_\alpha{}^\beta\| : U \to SO(2). \tag{132}$$

To solve the Euler-Lagrange equations we choose a particular $\mathbf{a}$-orthonormal moving frame defined by eqs.(27) and (131) with

$$e_a{}^\kappa = e^\varphi \delta_a^\kappa, \quad e^a{}_\kappa = e^{-\varphi}\delta_\kappa^a, \tag{133}$$

and we will call, following [14], this choice of the moving frame a *conformal gauge*.

**Theorem 3** [14] *For any solution of the Euler-Lagrange equations (121) in the conformal gauge there exists a scalar function f such that the SO(2)-connection of eq.(112) has the form*

$$\omega^{\alpha} = e^{\alpha\beta}\partial_{\beta}(\varphi + f), \qquad (134)$$

*where two functions φ and f, considered as functions of the complex variable z defined by eqs.(126) and (127), satisfy the following system of equations:*

$$4f_{z\bar{z}} + (f^2 - \Lambda)e^{-2\varphi} = 0,$$
$$4\varphi_{z\bar{z}} + (f^2 + f - \Lambda)e^{-2\varphi} = 0, \qquad (135)$$

*and*

$$f_{zz} + f_z^2 + 2\varphi_z f_z = 0,$$
$$f_{\bar{z}\bar{z}} + f_{\bar{z}}^2 + 2\varphi_{\bar{z}} f_{\bar{z}} = 0. \qquad (136)$$

*Inversely, for two functions φ and f satisfying eqs.(135) and (136), the zweibein and the SO(2)-connection constructed using formulas (133) and (134) satisfy the Euler-Lagrange equations (121).*

The general solution of the system of equations (135) and (136) is given by the following theorem.

**Theorem 4** [14] *Let D be an arbitrary connected two-dimensional domain on the complex plane $\mathbb{C}$, $f, \varphi \in C^2(D)$ and for $z \in D$ the partial derivative $f_z$ or $f_{\bar{z}}$ either equals identically zero or everywhere differs from zero. Next, let us denote by $w: D \to \mathbb{C}$ the complex coordinate function defined by the differential equation*

$$w'(\bar{z}) = e^{\overline{w(z)}} \neq 0 \qquad (137)$$

*and being an arbitrary holomorphic nonconstant function defined in some domain W, $D \subseteq W \cap \overline{W}$. Then any of the system of equations (135) and (136) in D belongs to one of two classes:*
(i) $f_z = 0$ or $f_{\bar{z}} = 0$,

$$f = \pm\sqrt{\Lambda} \qquad (138)$$

*and*

$$e^{-2\varphi} = \frac{w'\overline{w}'}{(aw\overline{w} + bw + \overline{b}\overline{w} + d)^2}, \qquad (139)$$

*where two real constants a, d, and one complex constant b satisfy the condition*

$$ad - b\bar{b} = \mp\frac{\sqrt{\Lambda}}{4}. \qquad (140)$$

*One must choose either upper or lower signs in eqs.(138) and (140).*
(ii) $f_z \neq 0$ or $f_{\bar{z}} \neq 0$,

$$f = h(w + \bar{w}),  \tag{141}$$

and

$$e^{-2\varphi} = h'w'\bar{w}'e^h, \qquad h' > 0,  \tag{142}$$

where *h is the real valued one real argument function defined by the ordinary differential equation*

$$4h' = -\left[\left(h^2 - 2h + 2 - \Lambda\right)e^h + A\right], \qquad A = \text{const.},  \tag{143}$$

where *A is an arbitrary constant.*

It follows from the above theorem that in the case (ii) we have [14]

$$R = -\frac{\mu}{\sigma}h, \qquad T^2 = 4\frac{\mu}{\sigma}h'e^{-h},  \tag{144}$$

while the case (i) is described by eq.(129), that is:

$$R = R_0 = \pm\frac{\mu}{\sigma}\sqrt{\Lambda}, \qquad T = 0.  \tag{145}$$

We see that if the influence of dislocations is neglected, then the solution admits scalar curvatures of both signs and the scalar curvature *R* covers with the scalar curvature of the Riemannian material space $M_a = (M, \mathbf{a})$. Let us compute the value of the action (118) for the constant curvature Riemannian material space [14]. Substitution of the solution (145) yields

$$\mathbb{S}_0 = \int_M L(R_0, 0)dF = \frac{1}{2}\sigma R_0^2 F(M)\chi(M),  \tag{146}$$

where *F(M)* is the surface area of the graphene sheet. Because the graphene sheets under consideration are homeomorphic to the disk so their Euler characteristic $\chi(M)$ equals to 1 and therefore, taking into account the relation between the Euler characteristic, the scalar curvature, and the area, we obtain that should be [14]

$$R_0 = \frac{4\pi}{F(M)}.  \tag{147}$$

It means that, for the constant curvature graphene sheets, admissible are only *positive Gauss curvatures*. Thus, taking into account [14], our assumption concerning material spaces of graphene sheets and "Theorem of Embeddings" quoted in [39] (Section 6), we conclude:

**Conclusion 3** *Let $M_a = (M, \mathbf{a})$ be a smooth oriented complete constant curvature Riemannian material space homeomorphic to the disk and with the finite surface area F(M). If the influence of dislocations is neglected (T = 0), then field equations (121) defines the Riemannian material space of constant positive Gauss curvature $K = R_0/2$ (in designations of [32]) admitting a smooth global isometric embedding into the orthogonal configurational space of graphene sheets. In the limit $F(M)\to\infty$, the range of this embedding is a developable surface.*

The case (ii) of Theorem 4 enables to formulate a rule according to which the torsion-type defects influence the material Riemannian metric **a**:

**Conclusion 4** *It follows from eqs.(119),(126), (130), (137), and (142) that the following formula holds:*

$$e^{-2\varphi} = \left(\frac{Tl_0}{2\sqrt{2}}\right)^2 e^{2h} w' \overline{w}'. \qquad (148)$$

*If additionally Statement 4 is taken into account, then we can take $l_0 = l(\theta_0)$ and the above formula defines, in a reference effective temperature $\theta_0$ of the graphene sheet, the influence of dislocations on the material Riemann metric* **a.**

Let us consider the canonically parametrized $\nabla^a$- geodesic corresponding to the conformal gauge and with the natural parameter $\tau$, $[\tau]$ = cm (see Section 4). If the manifold is equipped with the metric given by eq.(130) and by part (ii) of Theorem 4, then eq.(61) is equivalent to one complex equation [14]:

$$\frac{d^2 z}{d\tau^2} = 2 \frac{\partial \varphi}{\partial z}\left(\frac{dz}{d\tau}\right)^2. \qquad (149)$$

This equation can be integrated and assuming that $dz/d\tau \neq 0$ (otherwise the geodesic will be a point), we get the first integral $c_a$ [14]

$$\left|\frac{dz}{d\tau}\right| = c_a e^{\varphi}, \quad c_a > 0, \quad [c_a] = \text{cm}^{-1}, \qquad (150)$$

where it was denoted $|z| = \sqrt{z\overline{z}} = \sqrt{x^2 + y^2}$ and $c_a$ is a constant. The dimensionless length element $d\xi$ of the **a-**geodesic is given by:

$$\begin{aligned} d\xi^2 &= e^{-2\varphi} dz d\overline{z} = c_a^2 d\tau^2, \\ \xi &= l/l_0, \quad [l] = [l_0] = \text{cm}, \end{aligned} \qquad (151)$$

where $l_0$ is the characteristic length defined by eq.(126). Notice that the natural parameter $s$, $[s]$ = cm, of the canonically parametrized $\nabla$-geodesic of eq.(61) can be chosen, according to eqs.(65), (66) and (70), in the form defined by eq.(62).

## 7. Conclusions and remarks

While the methods of description of continuously defective three-dimensional crystal structures (mainly – Bravais crystals) are well known, the knowledge how to describe the continuously defective graphene sheets is not on the satisfactory level yet. Let us remind, in order to facilitate the formulation of conclusions and remarks concerning this topic, some facts mentioned already in the paper.

First of all, the problem already appears on the level of the description of the *continuized crystalline structure* (Section 2) without defects. Namely, in the case of corrugated graphene sheets we are dealing, in contrast to the continuously defective three-dimensional crystal structures, with the solid body which does not possess an undistorted spatial configuration (Section 4).

Important differences exist even in the case of dislocations and disclinations being defects common for two and three-dimensional crystal structures (Section 1). Dislocations and dis-clinations are *line defects* of a crystal lattice which define the distortions of its translational and rotational symmetries, respectively. These distortions are quantitatively characterized by the so called Burgers and Frank vectors (Section 1). Both the Burgers vector and its local continuous counterpart (Section 3) define the translational distortion and can be defined in terms of a two-dimensional graphene sheet. Moreover, in graphene sheets, only edge dislocations exist. The Frank vector (Section 1) is a rotation vector and thus cannot be defined in terms of the two-dimensional geometry. However the measure of rotational distortion called the Frank angle is still the well-defined quantity. Further, according to the topological classification of line defects, while in the three-dimensional case there exists exactly one type of irremovable distortion corresponding to the defects of rotation type, in the two-dimensional case, the rotation type defects with irremovable non-equivalent distortions are in one-to-one correspondence with the non-zero integers (Section 1).

There are also defects of crystalline structure specific for graphene. For example, topological defects in graphene can be formed by replacing a hexagon by n-sides polygon. A *pentagon* induces positive curvature while a *heptagon* induces the negative curvature and dislocations as well as disclinations may be formed as the combination of these curvature-type topological defects. (Section 1)

However, the *curvature-type topological defects* can be treated as elementary topological defects only in the discrete description of defective crystal structures. It is not the case of continuized description of these structures. Namely (see Section 2), we are considering here *classical crystals*, that is, crystals possessing as local degrees of freedom the translational degrees of freedom only. Consequently, this limiting process is possible for dislocations but not for disclinations. It follows from the fact that disclinations are specified by a finite angle (rotational symmetry of the crystal). This means that a crystal with disclinations cannot be continuized in Euclidean, or, more generally, in flat space (space without curvature). If so, the only *elementary line defects* in flat classical continuized crystals are dislocations (or edge dislocations only if we are dealing with graphene sheets – Sections 1 and 2). Disclinations are, in the continuous limit, rather a type of continuous distribution of dislocations than a separate kind of line defects; (Sections 5 and 6; see also Hypothesis 1 – Section 2).

Because the corrugated graphene sheets considered in the paper are homeomorphic to developable surfaces, dislocation lines in these sheets can be considered, in the *continuous limit approximation*, as Volterra-type *effective edge dislocation lines* (Sections 1 and 2) which are defined as smooth curves located in the such sheet and represented by their embedding in the material Riemannian space (Section 3). This representation is defined by a local (but not infinitesimal – see Section 2) counterpart of the Burgers vector called *local Burgers vector*, by the *scalar density of dislocations*, and by a congruence of virtual effective edge dislocation lines (Definitions 1 and 2 in Section 3). The dimensional analysis of the geometrical relations describing the continuous distributions of dislocations shows that this scalar density of dislocations ought to be defined as the density of the total number of these line defects located in the graphene sheet (Sections 2 and 3).

The material structure of continuously defective corrugated graphene sheets is modelled by a *material space* defined as the Riemannian material space additionally endowed with a certain covariant derivative (Section 2 and remarks at the beginning of the Section 6). Particularly, *continuously dislocated* corrugated graphene sheets are described with the help of the Riemann-Cartan geometry (Sections 5 and 6). It is shown (Conclusion 1) that if a continuously defective corrugated graphene sheet has the flat Riemann-Cartan material space, then dislocations are the only source of corrugations of this graphene sheet. In this particular case we are dealing with *secondary curvature-type defects* created by the distribution of dislocations. The case of continuously defective corrugated graphene sheets with secondary curvature-type defects is described in details in Conclusions 1, 2 and Statement 3. Particularly, the compatibility condition with the isothermal geometry [39] of these sheets is formulated. Moreover, it is stated that the Gaussian curvature of the Riemannian material space of these sheets can be interpreted as a "charge density" of their torsion-type defects.

We conclude that, in contrast to the discrete case, the continuously defective corrugated graphene sheets reveal two different kinds of defects only: the *curvature-type defects* and the *torsion-type defects* (Section 6). It ought to be stressed that while the torsion-type defects can be represented by the effective edge-type line defects (Section 3), the curvature-type defects have not a representation in the form of "effective disclinations". Moreover, because the considered graphene sheets are homeomorphic to developable surfaces ([39], Section 6), the torsion-type defects can be treated as elementary defects of continuously defective corrugated graphene sheets while the curvature-type defects can have the secondary character (Section 5; see Hypothesis 2). However, the proposed material geometry of these sheets admits also more general curvature-type defects (Sections 2 – 6).

In Conclusion 4 is given a formula which describes, basing oneself on the *variational field equations* of the material Riemann-Cartan geometry formulated in a reference effective temperature of the corrugated graphene sheet [39], the influence of torsion-type defects (that is, edge dislocations in our case) on the metric of the material space. If the influence of dislocations is neglected, this variational geometry defines the Riemannian material space of constant positive Gauss curvature admitting a smooth global isometric embedding into the orthogonal configurational space of graphene sheets (Conclusion 3). The proposed variational geometry of the material space is closely related with the existence of a characteristic length parameter of the continuously defective corrugated graphene sheet (Section 6; see also Section 2 - remarks concerning the notion of continuized crystals).

**Appendix – Torsion and curvature**

Let $\nabla$ be a covariant derivative on a differential manifold $M$, $\mathrm{W}(M)$ – the module of smooth vector fields on $M$ tangent to $M$, $\mathrm{W}(M)^*$ - the module of smooth covector fields (1-forms) on $M$ ([34]; see also e.g.[40] – Appendix), $\mathbf{u}, \mathbf{v} \in \mathrm{W}(M)$, $\mathrm{w} \in \mathrm{W}(M)^*$, $e = (\mathbf{e}_a)$ - a base of the module $\mathrm{W}(M)$ (called also a *moving frame* on $M$), $e^* = (\mathrm{e}^a)$ - the base the module $\mathrm{W}(M)^*$ dual to $e$ (called also the *moving coframe* dual to $e$). We define:

$$\nabla_\mathbf{v} \mathrm{w}(\mathbf{u}) = \partial_\mathbf{v}(\mathrm{w}(\mathbf{u})) - \mathrm{w}(\nabla_\mathbf{v} \mathbf{u}), \qquad \mathbf{u} = u^a \mathbf{e}_a, \qquad \mathbf{v} = v^a \mathbf{e}_a,$$
$$\nabla \mathbf{v} = \nabla_{\mathbf{e}_a} v^b \mathrm{e}^a \otimes \mathbf{e}_b, \qquad \nabla \mathbf{v}(\mathbf{u}) \equiv \nabla_\mathbf{u} \mathbf{v} = \left(u^a \nabla_{\mathbf{e}_a} v^b\right) \mathbf{e}_b.$$
(152)

The *connection forms* $\omega^a{}_b$, and the *Christoffel symbols* $\Gamma^a_{bc}$ of the covariant derivative $\nabla$ are defined by the rules:

$$\nabla \mathbf{e}_b = \omega^c{}_b \otimes \mathbf{e}_c, \quad \omega^c{}_b = \omega_a{}^c{}_b \mathbf{e}^a = \omega_\alpha{}^c{}_b du^\alpha, \quad \mathbf{e}^a = e^a{}_\alpha du^\alpha,$$
$$\nabla_{\mathbf{e}_a} \mathbf{e}_b = \Gamma^c_{ab} \mathbf{e}_c, \quad \omega_a{}^c{}_b \equiv \Gamma^c_{ab}, \quad \omega_\alpha{}^c{}_b = \omega_a{}^c{}_b e^a{}_\alpha. \tag{153}$$

Then

$$\omega^a{}_b(\mathbf{v}) = \mathbf{e}^a(\nabla_\mathbf{v} \mathbf{e}_b), \tag{154}$$

and

$$\Gamma^a_{bc} = \mathbf{e}^a(\nabla_{\mathbf{e}_b} \mathbf{e}_c) = -(\nabla_{\mathbf{e}_b} \mathbf{e}^a)(\mathbf{e}_c). \tag{155}$$

Notice that if the covariant differentiation calculus is formulated in terms of vector bundles, then the *Christoffel symbols* are usually called *connection coefficients*. However, in the paper we are dealing with the formulation of this calculus based on the notion of covariant derivatives in linear modules of fields defined on manifolds ([34] or e.g. [40] – Appendix), and thus the term of "Christoffel symbols" will be used even in the case of a nonmetric covariant derivative.

The *object of anholonomy* $C^c_{ab}$ of the moving frame $e$ is defined by means of the Lie bracket of the vector fields defining $e$:

$$[\mathbf{e}_a, \mathbf{e}_b] := \mathbf{e}_a \circ \mathbf{e}_b - \mathbf{e}_b \circ \mathbf{e}_a = C^c_{ab} \mathbf{e}_c, \tag{156}$$

where the base vector fields are identified with the linear differential operators and, correspondingly, their dual covectors are identified with the 1-forms:

$$\mathbf{e}_a = e_a{}^\alpha \partial_\alpha, \quad \mathbf{e}^a = e^a{}_\alpha du^\alpha,$$
$$\mathbf{e}^a(\mathbf{e}_b) = e^a{}_\alpha e_b{}^\alpha = \delta^a_b, \quad e^a{}_\alpha, e_b{}^\alpha \in C^\infty(M). \tag{157}$$

The so called *Jacobi identity* holds [34]:

$$C^a_{bc} = \mathbf{e}^a([\mathbf{e}_b, \mathbf{e}_c]) \in C^\infty(M), \quad C^a_{bc} = -C^a_{cb},$$
$$C^e_{ab} C^c_{de} + C^e_{bd} C^c_{ae} + C^e_{da} C^c_{be} + \partial_d C^c_{ab} + \partial_a C^c_{bd} + \partial_b C^c_{da} = 0, \tag{158}$$

where it was denoted $\partial_a \equiv \partial_{\mathbf{e}_a} \cong \mathbf{e}_a$ (see, e.g. [40] – Appendix). If $C^a_{bc} = 0$, then $e$ is called a *commutative base*.

The torsion tensor **S**, treated as a $C \equiv C^\infty(M)$-linear mapping $S \in L_C(W \times W; W)$, $W \equiv W(M)$, is defined by

$$T(\mathbf{u}, \mathbf{v}) \equiv 2S(\mathbf{u}, \mathbf{v}) = \nabla_\mathbf{u} \mathbf{v} - \nabla_\mathbf{v} \mathbf{u} - [\mathbf{u}, \mathbf{v}], \tag{159}$$

has the components $S^c_{ab}$ given by:

$$T^a_{bc} \equiv 2S^a_{bc} = \mathbf{e}^a(T(e_b, e_c)) = \Gamma^a_{bc} - \Gamma^a_{cb} - C^a_{bc}, \tag{160}$$

and can be written in the form:

$$\mathbf{S} = \tau^c \otimes \mathbf{e}_c, \quad \tau^c = d\mathbf{e}^c = S_{ab}{}^c \mathbf{e}^a \wedge \mathbf{e}^b,$$
$$\mathbf{e}^a \wedge \mathbf{e}^b = \mathbf{e}^a \otimes \mathbf{e}^b - \mathbf{e}^b \otimes \mathbf{e}^a. \tag{161}$$

The curvature tensor **R** is treated as a mapping $W \times W \to L_C(W; W)$ defined by the rules

$$R(\mathbf{u}, \mathbf{v}) \equiv R_{\mathbf{u},\mathbf{v}} = [\nabla_\mathbf{u}, \nabla_\mathbf{v}] - \nabla_{[\mathbf{u},\mathbf{v}]}, \quad R_{\mathbf{e}_a, \mathbf{e}_b} \mathbf{e}_c = R_{abc}{}^d \mathbf{e}_d, \tag{162}$$

where the components $R_{abc}{}^d$ have the form (in designations of [34]):

$$R_{abc}{}^d = \mathbf{e}^d\left(R_{\mathbf{e}_a,\mathbf{e}_b}\mathbf{e}_c\right) = \left(\partial_a \Gamma^d_{bc} - \partial_b \Gamma^d_{ac}\right) - \left(\Gamma^e_{ac}\Gamma^d_{be} - \Gamma^e_{bc}\Gamma^d_{ae}\right) - C^e_{ab}\Gamma^d_{ec}. \tag{163}$$

Let us consider a local coordinate system $(O, u)$, $O \subset M$, $u = (u^\alpha) : O \to \mathbb{R}^n$ - a smooth mapping, $[u^\alpha]$ = cm. In this coordinate system the components of the curvature tensor defined by eq.(163) cover with designations which are used e.g. in [9] and [32] but differ from those used e.g. in [6]. Namely, denoting by $R^\alpha{}_{\beta\mu\sigma}$ the designation of components in [6], we have:

$$R^\alpha{}_{\beta\mu\sigma} = R_{\mu\sigma\beta}{}^\alpha. \tag{164}$$

The designation of components such as in [6] is used e.g. in the papers [14], [43], [46], [47]. If the condition

$$\Gamma^a_{bc} = 0 \tag{165}$$

holds, then eq.(153) defines the so called *teleparallelism* on $M$ defined by a base $e = (\mathbf{e}_a; a = 1,\ldots,n)$ (see e.g. [32], [34]) and we will denote the corresponding uniquely defined covariant derivative by $\nabla^e$. If additionally the condition

$$\nabla^e \tau^a = 0, \quad a = 1, 2, \ldots, n, \tag{166}$$

is fulfilled, then the teleparallelism is called *closed* [37]. It is equivalent to the condition that the moving frame $e$ spans a real Lie algebra $g$ $e$-parallel vector fields:

$$g = \left\{\mathbf{v} = v^a \mathbf{e}_a : v^a = \text{const.}\right\}. \tag{167}$$

The scalars $C^c_{ab}$ of eq.(156) are now the structure constants of this Lie algebra. The $\nabla^e$ - Christoffel symbols $\Gamma^{e,\gamma}_{\alpha\beta}$ have, in a local coordinate system $u = (u^\alpha)$ on $M$, the form

$$\Gamma^{e,\gamma}_{\alpha\beta} = e_c{}^\gamma \partial_\alpha e^c{}_\beta \tag{168}$$

and the components of torsion tensor take the following form:

$$S_{\alpha\beta}{}^\gamma = \Gamma^{e,\gamma}_{[\alpha\beta]} = e_c{}^\gamma \tau^c{}_{\alpha\beta},$$
$$\tau^c = \tau^c{}_{\alpha\beta} du^\alpha \wedge du^\beta, \qquad \tau^c{}_{\alpha\beta} = \partial_{[\alpha} e^c{}_{\beta]}, \tag{169}$$

where eqs.(157) and (161) were taken into account.

### Acknowledgements

This paper contains results obtained within the framework of the research project N N501 049540 financed from Scientific Research Support Fund in 2011-2014. The author is greatly indebted to the Polish Ministry of Science and Higher Education for this financial support.